\documentclass[bibyear]{aa}  
\usepackage{graphicx,url}
\usepackage[varg]{txfonts}     
\usepackage{gensymb} 
\usepackage[breaklinks]{hyperref}  
\usepackage{pdfcomment,acronym}      
\usepackage{color}
\definecolor{darkblue}{rgb}{0,0,1}
\hypersetup{
  colorlinks=true,   
  urlcolor=blue,     
  linkcolor=red,     
  citecolor=darkblue 
}
\usepackage{natbib,twoopt}
\bibpunct{(}{)}{;}{a}{}{,}    

\makeatletter
\newcommand{\bibnote}[2]{\@namedef{#1note}{#2}}
\newcommand{\biblink}[2]{\@namedef{#1link}{#2}}
\makeatother

\makeatletter
\newcommandtwoopt{\citeads}[3][][]{\href{http://adsabs.harvard.edu/abs/#3}%
{\def\hyper@linkstart##1##2{}%
\let\hyper@linkend\@empty\citealp[#1][#2]{#3}}\biblink{#3}{\href{http://adsabs.harvard.edu/abs/#3}{ADS}}}
\newcommandtwoopt{\citepads}[3][][]{\href{http://adsabs.harvard.edu/abs/#3}%
{\def\hyper@linkstart##1##2{}%
\let\hyper@linkend\@empty\citep[#1][#2]{#3}}\biblink{#3}{\href{http://adsabs.harvard.edu/abs/#3}{ADS}}}
\newcommandtwoopt{\citetads}[3][][]{\href{http://adsabs.harvard.edu/abs/#3}%
{\def\hyper@linkstart##1##2{}%
\let\hyper@linkend\@empty\citet[#1][#2]{#3}}\biblink{#3}{\href{http://adsabs.harvard.edu/abs/#3}{ADS}}}
\newcommandtwoopt{\citeyearads}[3][][]%
{\href{http://adsabs.harvard.edu/abs/#3}
{\def\hyper@linkstart##1##2{}%
\let\hyper@linkend\@empty\citeyear[#1][#2]{#3}}\biblink{#3}{\href{http://adsabs.harvard.edu/abs/#3}{ADS}}}
\makeatother

\newacro{ADS}{Astrophysics Data System}
\newacro{NLTE}{non-local thermodynamic equilibrium}
\newacro{NASA}{National Aeronautics and Space Administration}

\begin{document}


\title{Study of the spatial association between an active region jet and a nonthermal type~${\rm III}$ radio burst}
\subtitle{}

\author{
Sargam M. Mulay$^{1,2}$
\and
Rohit Sharma$^{3}$
\and
Gherardo Valori$^{4}$
\and
Alberto M. V\'{a}squez$^{5}$
\and
Giulio Del Zanna$^{2}$
\and
Helen Mason$^{2}$
\and
Divya Oberoi$^{6}$
}

\institute{Inter-University Centre for Astronomy and Astrophysics (IUCAA), Post Bag-4, Ganeshkhind, Pune 411007, India \\
\email{sargam@iucaa.in}
\and
DAMTP, Centre for Mathematical Sciences, University of Cambridge,
Wilberforce Road, Cambridge, CB3 0WA, UK\\
\and
University of Applied Sciences and Arts Northwestern Switzerland, Bahnhofstrasse 6, 5210 Windisch, Switzerland \\
\and
University College London, Mullard Space Science Laboratory, Holmbury St. Mary, Dorking, Surrey, RH5 6NT, UK \\
\and
Instituto de Astronom\'{i}a y F\'{i}sica del Espacio (IAFE), CONICET- UBA, Argentina\\
\and
National Centre for Radio Astrophysics (NCRA), Tata Institute of Fundamental Research, Savitribai Phule Pune University Campus, Pune, Maharashtra, India\\
}

\date{Received ; accepted }

 \abstract
{}
{We aim to investigate the spatial location of the source of an active region (AR) jet and its relation with associated nonthermal type~III radio emission.}
{An emission measure (EM) method was used to study the thermodynamic nature of the AR jet. The nonthermal type~{\rm III} radio burst observed at meterwavelength was studied using the Murchison Widefield Array (MWA) radio imaging and spectroscopic data. The local configuration of the magnetic field and the connectivity of the source region of the jet with open magnetic field structures was studied using a nonlinear force-free field (NLFFF) extrapolation and potential field source surface (PFSS) extrapolation respectively.} 
{The plane-of-sky velocity of the AR jet was found to be $\sim$136~km/s. The EM analysis confirmed the presence of low temperature 2~MK plasma for the spire, whereas hot plasma, between 5-8 MK, was present at the footpoint region which also showed the presence of Fe~{\sc xviii} emission. A lower limit on the electron number density was found to be 1.4$\times$10$^{8}$ cm$^{-3}$ for the spire and 2.2$\times$10$^{8}$~cm$^{-3}$ for the footpoint. A temporal and spatial correlation between the AR jet and nonthermal type III burst confirmed the presence of open magnetic fields. An NLFFF extrapolation showed that the photospheric footpoints of the null point were anchored at the location of the source brightening of the jet. The spatial location of the radio sources suggests an association with the extrapolated closed and open magnetic fields although strong propagation effects are also present.} 
{The multi-scale analysis of the field at local, AR, and solar scales confirms the interlink between different flux bundles involved in the generation of the type III radio signal with flux transferred from a small coronal hole to the periphery of the sunspot via null point reconnection with an emerging structure.}

\titlerunning{Type~{\rm III} radio emission and active region jets}
\authorrunning{Sargam M. Mulay et al.} 

\maketitle


\section{Introduction} \label{introduction}


At meter wavelengths, the large angular scale solar emission comprises emission from the solar disk which is produced by free-free thermal bremsstrahlung radiation. The dominant compact radio sources are mostly associated with active region locations. These radio sources are produced by the electrons trapped in the coronal loops. During the bursts, these trapped electrons are accelerated and escape via open magnetic field lines. In this process, these accelerated electrons emit radio emission (known as type~III radio burst) via the plasma emission mechanism dominantly occurring at the local fundamental frequency $\omega_p$, and/or at harmonics where $\omega_p \propto \frac{\sqrt{n_e}}{T_e^{3/2}}$, where $n_e$ and $T_e$ are the local plasma electron number density and temperature respectively \citep{Ginzburg58, Reiner92, Reid14, Cairns18, Krupar18, Krasnoselskikh18}. However, the observational signatures of the open magnetic field lines are hard to detect. Using radio imaging, the position of the radio sources can be deduced with respect to the background EUV event location. 

The spatial correlation of the flares and X-ray bursts as well as type~{\rm III} radio bursts has been known for some time \citep{Aschwanden95, Bastian98, Arzner05, Chen13, Reid17} and it provides a unique diagnostic for the nonthermal electrons (up to near-relativistic speeds), which are released into the heliosphere. This correlation may become weak with decreasing radio frequencies (specifically at meter waves). The major reasons for this could be magnetic field evolution, which appears both structurally and temporally, in addition to propagation effects, such as coronal scattering, refraction, and ionospheric effects \citep{Hollweg68, Cairns95}. However, type~{\rm III} diagnostics at various frequencies allow us to probe the emission at coronal locations beyond that of the triggering site. Solar active regions also frequently produce type~{\rm III} emission associated with EUV jets, that is, jets that are observed in EUV images of the low corona such as SDO/AIA. Recently, \citet{Klassen11, Innes11, Mulay16}, and \citet{Mohan19} show a temporal correlation of type~{\rm III} emission and EUV jets, while \citet{Chen13} show the locations of the radio sources at 1-2~GHz in the context of a jet eruption. 

\cite{Mulay16} carried out a multiwavelength study of twenty active region jets and found that 85\% of jets were temporally associated with nonthermal type~{\rm III} radio bursts (in the frequency range from 13~MHz to 20~kHz). The spatial correlation between the EUV jets and type~{\rm III} bursts was studied by extrapolating the photospheric magnetic field using the potential field source surface (PFSS) technique. The observations were interpreted in terms of nonthermal electrons, which accelerated during the reconnection outflow that was produced during the event, and then they escaped into the heliosphere along open field lines. Indeed, PFSS modeling always shows the presence of open field lines near the source region of the EUV jets that are close to the sunspots. This led the authors to conclude that AR jets are temporally and spatially associated with nonthermal type~III radio emission.

\begin{figure*}[!hbtp]
\begin{center}
\includegraphics[trim = 0cm 0cm 1.3cm 5.8cm,width=1.0\textwidth]{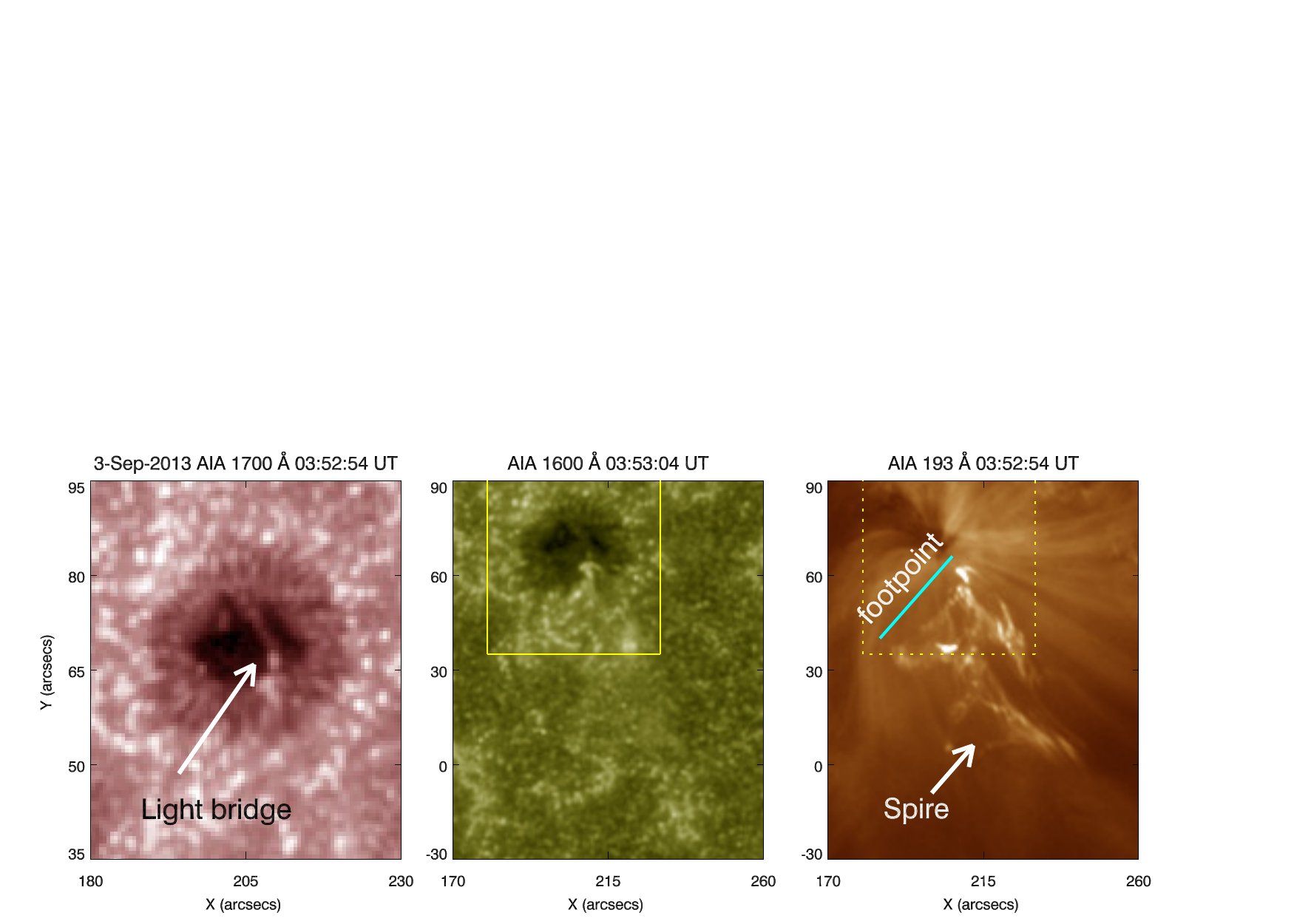}  
\caption{AIA images of a jet and a sunspot. Left panel: AIA 1700~{\AA} image of sunspot associated to active region NOAA 11836 (N11 W10). The light bridge is indicated by a white arrow. Middle and right panel: AIA 1600 and 193~{\AA} images during jet activity. The field-of-view shown in the left panel is denoted by a solid-line and dashed-line box in the middle and right panels, respectively. The extent of the footpoint region is shown by a cyan line and the spire region is shown by a white arrow.} \label{fig1_aia_context_img}
\end{center}
\end{figure*}


The fact that active region jets connect the source region that is next to a sunspot to the heliosphere is now of particular relevance for one of the main scientific goals of the Solar Orbiter \citep{Muller13}, which is to link the in-situ measurements of the solar wind to the source region where the solar wind originated. Such linking is notoriously difficult to establish, but having clear signatures, such as jets, would really help to make the connection unambiguous. On the other hand, we note that magnetic field extrapolations always carry with them some uncertainties related to the modeling, the resolution of the photospheric magnetic field, etc. Clearly, more reliable information on the source location of the type~{\rm III} bursts would be obtained with simultaneous imaging and spectral observations at radio wavelengths.





Unfortunately, radio imaging telescopes are not available for the frequency range from 13~MHz to 20~kHz which covers the interplanetary type~{\rm III} emission studied by \citet{Mulay16}. However, imaging and frequency observations are available at higher frequencies, which are formed in the lower corona, around one solar radii or below. Active regions typically emit during their lifetimes, the so-called radio noise storms, and have a persistent broad-band emission in the 100--500 MHz range, which is thought to be associated with  nonthermal electrons that continuously accelerate to relatively low energies. \citet{DelZanna11} studied such events observed with the Nancay radioheliograph (NRH) at frequencies between 150~MHz and 432~MHz, and found a spatial correlation between their location and the places where interchange reconnection would be expected to take place, on the basis of magnetic field modeling. \citet{DelZanna11} propose a unified physical interpretation of these noise storms and the so-called coronal outflows, which are open-field regions associated with sunspots where coronal lines form above 1~MK and show moderate outflows (10--20~km/s).

\begin{figure*}[!hbtp]
\begin{center}
\includegraphics[trim = 1cm 5cm 2cm 0cm,width=0.7\textwidth]{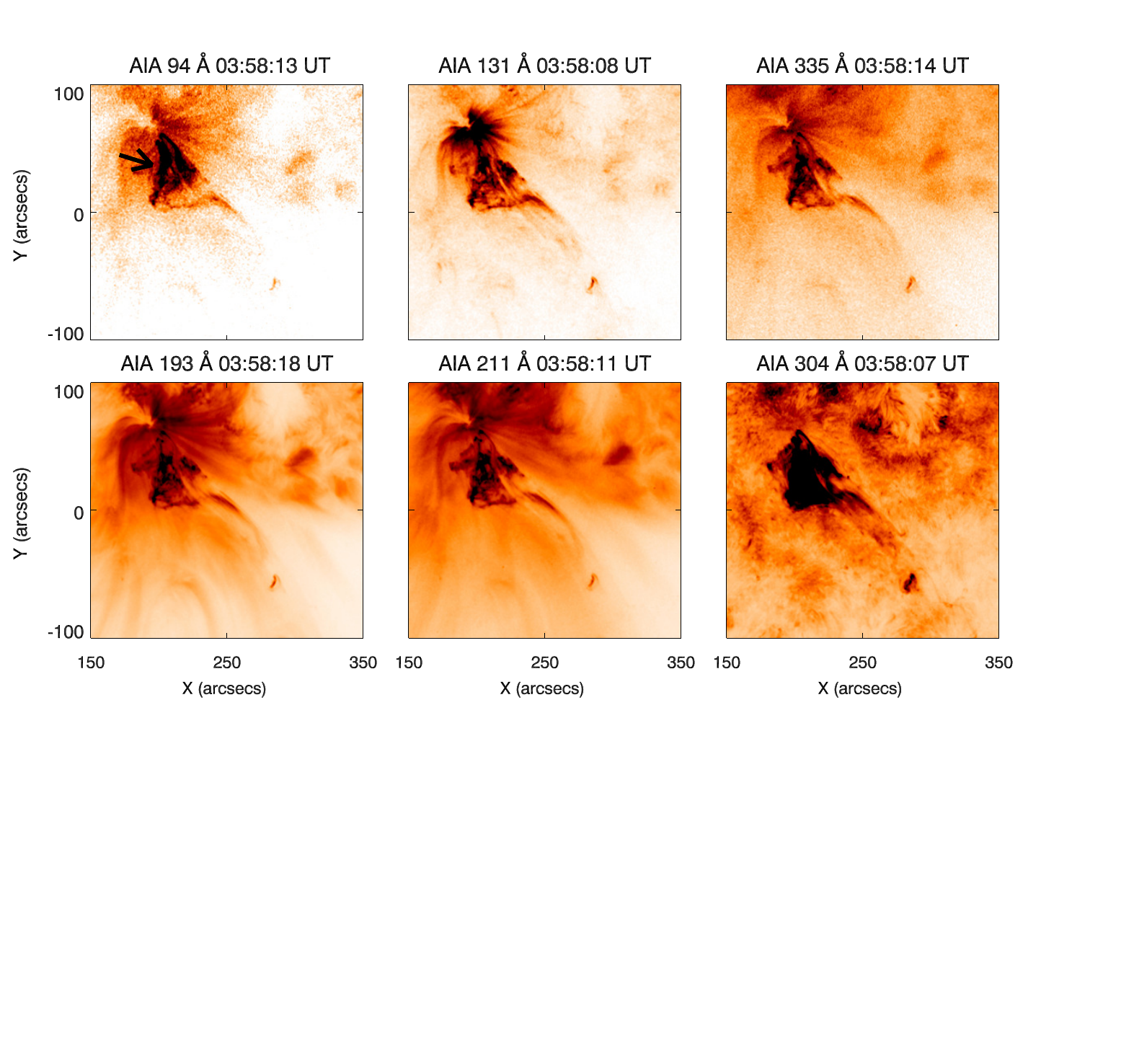}  
\caption{Ejection of AR jet in six AIA channels (reverse color images). The black arrow in the AIA 94~{\AA} channel indicates a small loop structure formed at the footpoint region during the evolution of the jet.} \label{fig2_jet_evolution}
\end{center}
\end{figure*}

At their low-frequency end, noise storms are often found to be accompanied by interplanetary type~{\rm III} emission. In the lower frequency range, the recently developed ground-based radio imaging instrument, Murchison Widefield Array \citep[MWA; ][]{Tingay13, Lonsdale09} is a suitable instrument for the study of type~{\rm III} emission. It is capable of simultaneously providing dynamic spectrum and radio imaging in the frequency range from 80~MHz to 300~MHz. MWA is designed to provide observations at higher coronal heights which can be further compared with interplanetary radio emission. 

After a careful search through the MWA data sets, we found a nonthermal type~{\rm III} radio burst, which is associated with an on-disk active region EUV jet (AR jet). The radio imaging helped to locate the radio sources in the lower corona during the jet activity and also helped to identify the source location for the interplanetary counterparts of type~III emission, which were detected by the WAVES instrument. A comprehensive analysis has been carried out to understand the spatial and temporal correlation between the AR jet and type~III burst.

This study provides an important opportunity to advance our understanding of plasma parameters, propagation of energetic particles in the solar atmosphere, and the role of photospheric and coronal magnetic activity in the ejection of active region jets. In this paper, the initiation of jet activity has been thoroughly studied using the line-of-sight and vector magnetic field data from the HMI instrument. The observed  structure of the jet was compared to that of the extrapolated coronal magnetic field as determined with two different methods, a nonlinear force-free field (NLFFF) extrapolation and a potential field source surface (PFSS) one. The study focused on understanding the dynamics involved in the photosphere as well as the changes and evolution of the magnetic structure at coronal heights.


The paper is structured as follows: UV and EUV imaging observations of the jet from AIA and radio observations from MWA are discussed in Sections~\ref{ch10_aia_obs} and \ref{type3_burst} respectively. The photospheric magnetic extrapolations are presented in Section~\ref{extrapolation}. We compare the extrapolated magnetic field with radio sources in Section~\ref{section5} and the results are summarized in Section~\ref{discussion}.


\section{UV and EUV imaging observations from AIA}  
\label{ch10_aia_obs}


The AIA instrument observed a series of homologous and recurrent AR jets on September~2 and 3, 2013. In this paper, we focus on one AR jet event observed on September~3, 2013 where we have simultaneous radio imaging and spectroscopic observations from MWA, and EUV and UV imaging from AIA. High-resolution (0.6$\arcsec$), high-cadence (12~sec) full disk AIA data was obtained and prepared by following the standard procedure (\texttt{aia\_prep.pro}). The AIA channels were co-aligned using the sunspot as a reference region.

Figure~\ref{fig1_aia_context_img} (left panel) shows a leading sunspot in the AIA 1700~{\AA} channel, which is associated with active region NOAA 11836 (N11~W10)\footnote{https://www.solarmonitor.org/?date=20130903}. The jet originated from the southern umbral-penumbral boundary of the leading sunspot. The activity was examined in all AIA channels before, during and after the jet evolution, that is, between 03:30 - 04:15~UT (See online Movie~1). Figure~\ref{fig1_aia_context_img} (middle and right panel) and Figure~\ref{fig2_jet_evolution} show a jet observed in different AIA channels. The jet activity started with a brightening at the footpoint at 03:48~UT in all AIA channels and the jet plasma disappeared by 04:03~UT. We recorded the lifetime of the jet as $\sim$15 min. During the evolution of the jet, a small loop structure formed at the footpoint region (shown by a small black arrow in the AIA~94~{\AA} channel in Fig.~\ref{fig2_jet_evolution}). One end of this loop is rooted in the umbral-penumbral boundary of the sunspot and in a region that is south of the sunspot. The jet originated from the southern edge of the light-bridge region and showed an untwisting motion of the spire (the same as was seen for the jet on August~10, 2010 as studied by \citealt{Mulay16}).

\begin{figure*}[!hbtp]
 \begin{center}
\includegraphics[trim=2.5cm 0.1cm 2.5cm 5cm,width=0.5\textwidth]{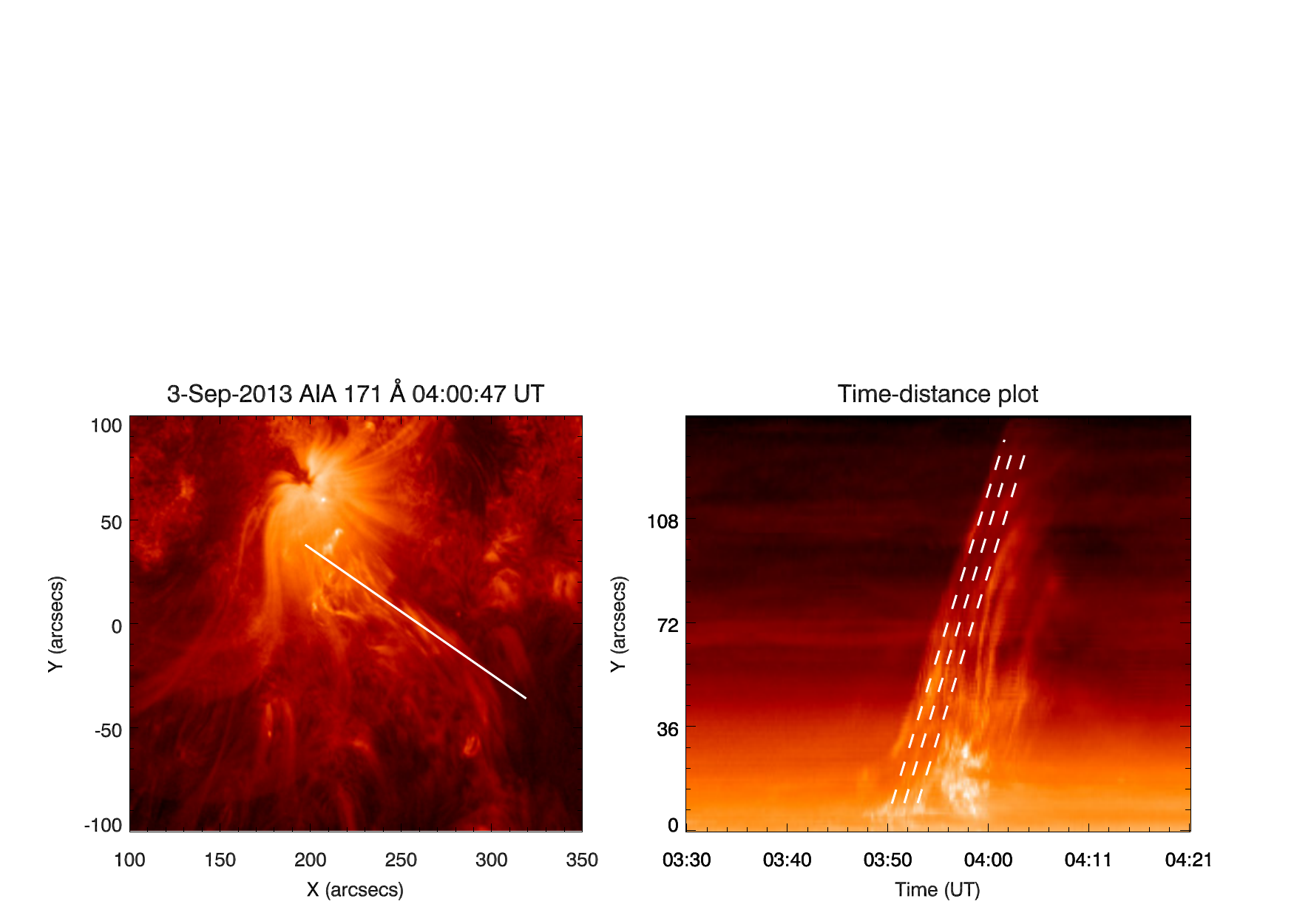}
\caption{Left panel: AIA 171~{\AA} image of jet at \mbox{04:00:47 UT}. The white overplotted line represents an artificial slit which was used to create a time-distance plot (the variation in the intensities along the artificial slit are shown as a function of time). Right panel: Time-distance plot for jet. The white dashed lines along the jet-spire structure are used for the plane-of-sky velocity calculation. The velocity was found to be \mbox{$\sim$136~km/s}.}
\label{fig3_aia_xt_plots}
\end{center}
\end{figure*}

The images in the AIA 171 (see Fig.~\ref{fig3_aia_xt_plots}, left panel), 193, and 211~{\AA} (see Fig.~\ref{fig2_jet_evolution}) channels show that the sunspot region was the source of a myriad of loops that obscured the footpoint region of the jet. Sunspot waves were observed to propagate along the EUV loops. They might affect the initiation mechanism of the jet, as discussed for another recurrent jet, as studied by \cite{Chandra15}.

The AIA 304~${\AA}$ channel is multi-thermal but on-disk dominated by cool He~{\sc II} emission. Having a peak sensitivity at log~\textit{T}~[K] = 4.7, the AIA 304~${\AA}$ channel shown in the bottom panel of Fig.~\ref{fig2_jet_evolution} does not detect the $\approx 1-2\,\rm{MK}$ EUV loops anchored in the sunspot. A brightening at the footpoint and untwisting motion of the spire is also seen in AIA 304~{\AA} images. During the evolution of the jet, the Geostationary Operational Environmental Satellite (GOES) did not observe any significant enhancement in X-ray flux, such as a solar flare.


\subsection{Plane-of-sky velocities using AIA 171~{\AA} images} 
\label{plane_of_sky_velocity_cal}

High cadence (12~sec) images from the AIA 171~{\AA} channel (dominated
by the Fe~\textrm{IX} 171.07~{\AA} line formed at log~\textit{T}~[K] =
5.85) during the evolution of the jet were used for a time-distance analysis. Figure~\ref{fig3_aia_xt_plots} shows the jet (left panel) observed at 04:00~UT in the AIA~171~{\AA} channel. The white line shows an artificial slit positioned along the jet-spire, which was used to measure the variation in the intensities as a function of time (right panel). The time-distance plot shows that the spire of the jet consisted of multiple thin strands. We calculated an average plane-of-sky velocity to be \mbox{$\sim$136 km/s} along white dashed lines. The result is found to be consistent with the range of values reported by \cite{Mulay16}.

\subsection{Emission measure analysis} 
\label{em_cal}

\begin{figure*}[!hbtp]
\begin{center}
\includegraphics[trim = 1cm 0cm 0cm 0cm,width=0.47\textwidth]{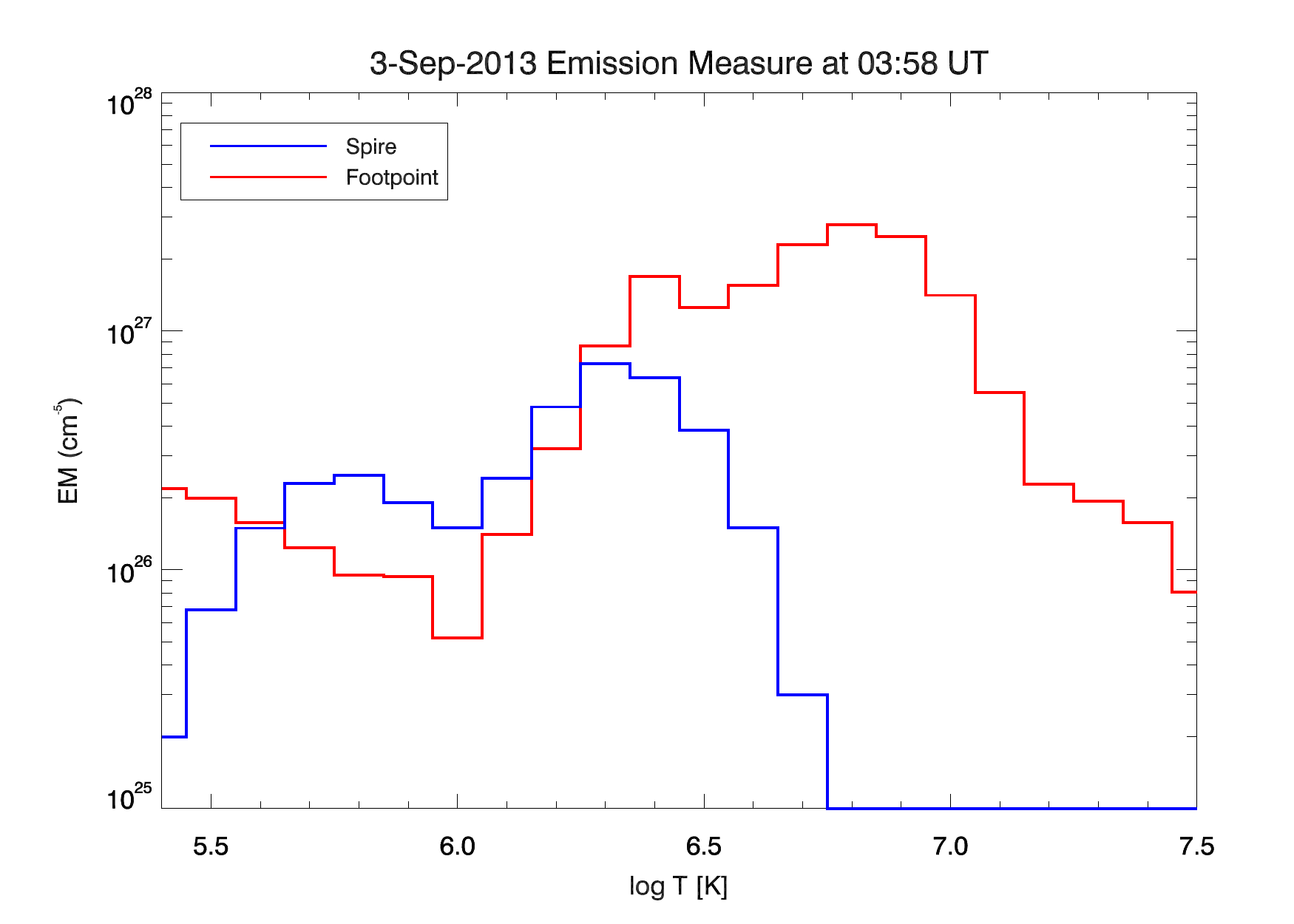}
\includegraphics[trim = 1.0cm 0cm 0cm 1cm,width=0.51\textwidth]{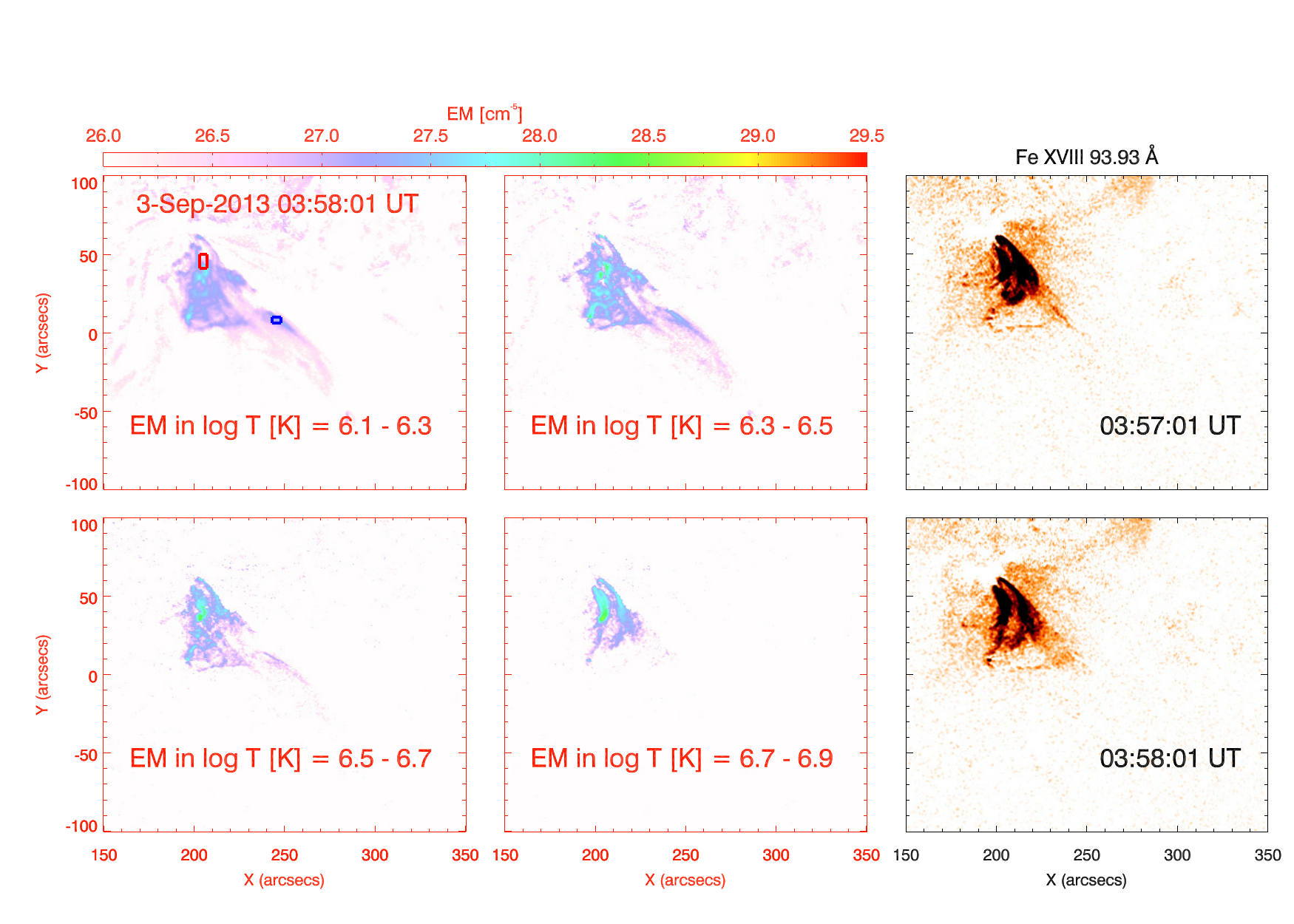}
\caption{Left panel: Emission measure curves for spire (blue) and footpoint (red) of jet. Right panel: Emission measure maps obtained for various temperature intervals (column~1 and 2) and Fe~{\sc xviii} 93.93~{\AA} (column 3, reverse color images) emission estimated from the AIA~94~{\AA} channel (see online Movie~1). The small blue and red boxes show the spire and footpoint regions, respectively, where the EM curves were created.} \label{fig_jet_em}
\end{center}
\end{figure*}

We performed an emission measure (EM) analysis using the inversion method developed by \cite{Cheung15} to study the thermal structure of the jet plasma. We coaligned AIA images in six channels (94, 131, 171, 193, 211, and 335~{\AA}) during the jet evolution at 03:58:01~UT. The method computes the temperature responses for six AIA channels (using the solarsoft routine \texttt{aia\_get\_response.pro}), using the CHIANTI atomic database \citep{Dere97} and coronal abundances (by combining the abundances from \cite{Feldman1992}, \cite{Grevesse1998}, and \cite{Landi2002}). Since most parts of the footpoint region were blocked by the sunspot loops, we subtracted the pre-jet image at 03:00~UT from the jet image at 03:58:01~UT. The sunspot loops were removed in the subtracted images, which provided a clear view of the jet structure. The EM analysis was carried out using these subtracted images. 

Figure~\ref{fig_jet_em} (right panel, column 1 and 2) shows EM maps at various temperature intervals. These EM maps show that multithermal plasma was present in the spire (log \textit{T} [K] = 6.1 - 6.5 (1.3-3 MK)) as well as in the footpoint (log \textit{T} [K] = 6.1 - 6.9 (1.3-8 MK)). Furthermore, we calculated EM curves (see Fig.~\ref{fig_jet_em}, left panel) for the spire and footpoint averaged over the blue and red boxed regions respectively. The EM curve for the spire (blue curve) showed two peaks; the first peak at log \textit{T} [K] = 5.8 (0.6 MK) and the second peak at log \textit{T} [K] = 6.3 (2~MK) (peak EM = 7.3$\times$10$^{26}$ cm$^{-5}$). The EM curve for the spire is not well constrained for log \textit{T} $<$ 5.8 and $>$ 6.4, and it falls sharply. The EM curve for the footpoint (red curve) peaked at log \textit{T} [K] = 6.8 (6.3~MK) (peak EM = 2.8$\times$10$^{27}$ cm$^{-5}$). The EM curve for the footpoint is not well constrained for log \textit{T} $<$ 6.3 and $>$ 6.9, hence it falls sharply and it also shows spurious enhancement of EM in the low (high) temperature range, that is, log \textit{T} $<$ 6.0 ($>$7.0). A small loop structure at the footpoint region (the same as seen in the AIA 94~{\AA} image of Fig.~\ref{fig2_jet_evolution}) is nicely observed in the EM maps created at log \textit{T} [K] = 6.7 to 6.9 (right panel, column 2, bottom panel). This result shows that the high temperature of 5-8~MK was present at the footpoint region. 

Furthermore, we investigated the reliability of high temperatures as seen in the footpoint region by estimating Fe~{\sc xviii} 93.932~{\AA} emission from the AIA 94~{\AA} channel. We used the empirical formula (I(Fe~{\sc xviii} (93.93~{\AA})) = I(94~{\AA}) - I(211~{\AA})/120 - I(171~{\AA})/450) given by \cite{DelZanna13b}. The method uses a combination of three AIA channels (94, 211, and 171~{\AA}). It removes the low temperature component from AIA 94~{\AA} images and provides Fe~{\sc xviii} 93.932~{\AA} emission. Figure~\ref{fig_jet_em} (right panel, column 3) shows the estimated Fe~{\sc xviii} 93.932~{\AA} emission in the jet at 03:57 (top panel) and 03:58~UT (bottom panel). The AR jet structure seen at 03:58~UT is very similar to that which was seen in the EM map created at log \textit{T} [K] = 6.7 - 6.9. Fe~{\sc xviii} 93.932~{\AA} is formed over a broad range of temperatures, and its maximum  abundance is at log \textit{T} [K] = 6.85. The analysis confirms the presence of high temperatures of log \textit{T} [K] = 6.8 (6.3~MK) in the footpoint region, which are consistent with the line formation (peak) temperature of Fe~{\sc xviii} 93.932~{\AA}. 

By assuming a cylindrical geometry, we measured the width of the spire as 3$\arcsec$ and 6$\arcsec$ for the footpoint from AIA images. We considered the column depth of the plasma along the line of sight to be the same as the widths and calculated a lower limit of the electron number density (\textit{N}$_{\rm e}$). We measured the total EM for the spire (5.4$<$log \textit{T}$<$6.8) and footpoint (5.4$<$log \textit{T}$<$7.5) (see Table~\ref{em_para_table}). By assuming a filling factor equal to unity, a lower limit of the electron number density (\textit{N}$_{\rm e}$) was found to be 1.4$\times$10$^{8}$~cm$^{-3}$ for the spire and 2.2$\times$10$^{8}$~cm$^{-3}$ for the footpoint. The density in the footpoint was found to be somewhat higher than the spire region.


 \begin{table}[!hbtp]
  \centering
 \caption{Physical parameters of AR jet.}
 \resizebox{9cm}{!} {
 \begin{tabular}{l c c}

 \hline
  
Physical parameters               &Spire &Footpoint\\

\hline


Peak log EM (cm$^{-5}$)  &26.9 &27.5  \\

Peak log \textit{T} [K]  &6.3  &6.8   \\

Total log EM (cm$^{-5}$)     &27.5 &28.2  \\

Electron number density (\textit{N}$_{\rm e}$) (cm$^{-3}$)    &1.4$\times$10$^{8}$ &2.2$\times$10$^{8}$           \\

\hline

  \end{tabular} \label{em_para_table}
}
\end{table}
\section{Nonthermal type~{\rm III} radio emission}  
\label{type3_burst}

\subsection{Radio spectrum analysis}  
\label{type3_spectra}

MWA took solar observations under proposal G0002 on September 3, 2013 from 03:40 to 04:40~UT and observed a group of nonthermal type~{\rm III} radio bursts in the lower corona from 03:48:16 to 04:04:48~UT. MWA observations have ten coarse frequency bands in the frequency range from 101~MHz to 298~MHz. Each coarse band has 64 fine spectral channels with a frequency resolution of 40~kHz and a temporal resolution of 0.5~sec. The interplanetary counterpart of the type~III burst was observed by the WAVES instrument\footnote{https://solar-radio.gsfc.nasa.gov/wind/index.html} \citep{Bougeret95} onboard the WIND spacecraft\footnote{https://wind.nasa.gov/} in the frequency range from 13~MHz to $\sim$200~kHz.

\begin{figure*}[!hbtp]
 \begin{center}
 \includegraphics[trim = 0cm 0cm 0cm 0cm,width=0.9\textwidth]{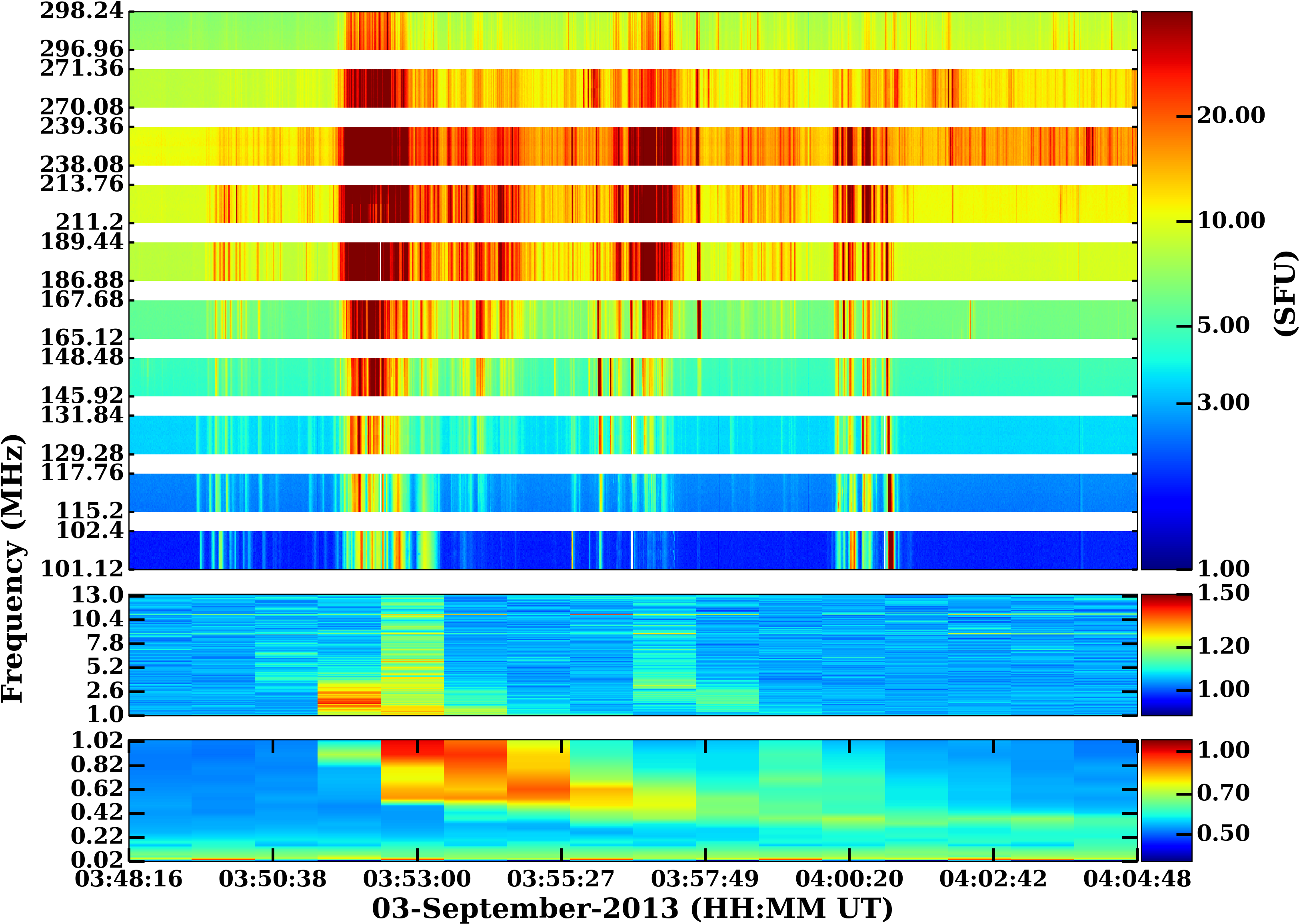} 
 \caption{Radio dynamic spectrum of nonthermal group of type~{\rm III} radio bursts observed on September~3, 2013 by MWA (top panel) and WAVES (middle and bottom panel). The colors in the MWA spectrum represent spectral flux density in solar flux unit (SFU) where 1~SFU = 10$^{4}$~Jansky = 10$^{-22}$ W/m$^{2}$Hz. The flux calibration has not been carried out for the WAVES spectrum, hence the colors show some arbitrary units of intensity.} \label{fig4_type3_spectrum}
 \end{center}
 \end{figure*}

Figure~\ref{fig4_type3_spectrum} shows half a second averaged radio dynamic spectrum from MWA (top panel) and a one-minute averaged radio spectrum obtained from WAVES (middle and bottom panel). The solar flux calibration for the MWA radio spectrum was performed using the baseline-based technique using a priori information about the well characterized MWA tiles and a sky model \citep{Oberoi17}. The MWA spectrum clearly shows that the type~III burst consists of some individual columns of energy releases, along with many fine
structures. The columns of type~III bursts show diverse values for
drifts. The peak energy release at 03:52~UT in the MWA spectrum continues in
the WAVES dynamic spectrum; this eventually leads to an interplanetary type III radio burst. The fine structures observed in the dynamic spectrum show a large variety in their time and frequency spans. Both lower coronal and interplanetary nonthermal type~{\rm III} emission were found to be temporally well correlated with the AR jet and provide evidence of particle acceleration during the jet ejection.

\subsection{Radio imaging analysis using MWA}  
\label{type3_imaging}

The spatial information for the type~{\rm III} radio burst was obtained by imaging the radio sources. The radio imaging at multiple frequencies facilitates the study of propagation as well as the movement of radio sources in the lower corona during the jet activity. For this study, three frequencies 101, 165, and 298~MHz were selected from MWA radio data. The extreme bands of 101 MHz and 298~MHz span a large range of coronal heights (see Table~\ref{Tab:centroid_table}), with 298~MHz being closest to, and 101~MHz being farthest from, the eruption site. The Common Astronomy Software Applications package\footnote{https://casa.nrao.edu/} (CASA) was used to process the MWA raw data. The standard radio analysis including calibration and deconvolution was carried out. 


The propagation effects through inhomogeneous and dynamic Earth's ionosphere posed a major challenge in determining the accurate source position at meterwaves. They can be seen as a shift in the positions of radio sources in the radio images \citep{Thejappa08}. In our analysis, a shift of a few arcmins in the solar position was observed. Therefore, a shift in the absolute location of the center of the Sun is present in the radio images with respect to the EUV solar images. This shift was corrected by computing the center of the solar disk using contour fitting to the solar disk in the radio images. 

\begin{figure*}[!hbtp]
 \begin{center}
 \includegraphics[trim = 0cm 0cm 0cm -1.5cm,width=0.30\textwidth]{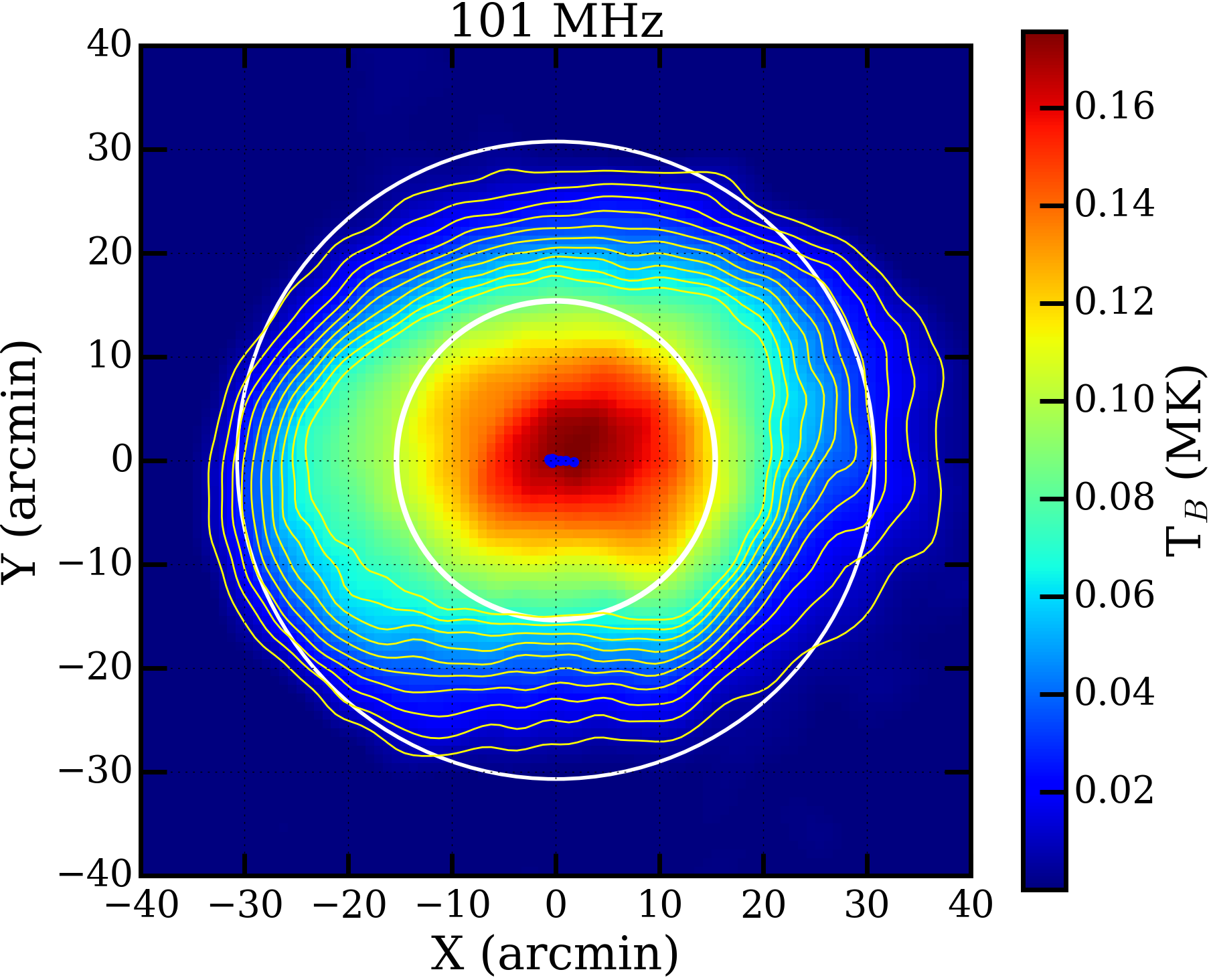}
 \includegraphics[trim = 0cm 0cm 0cm 0cm,width=0.30\textwidth]{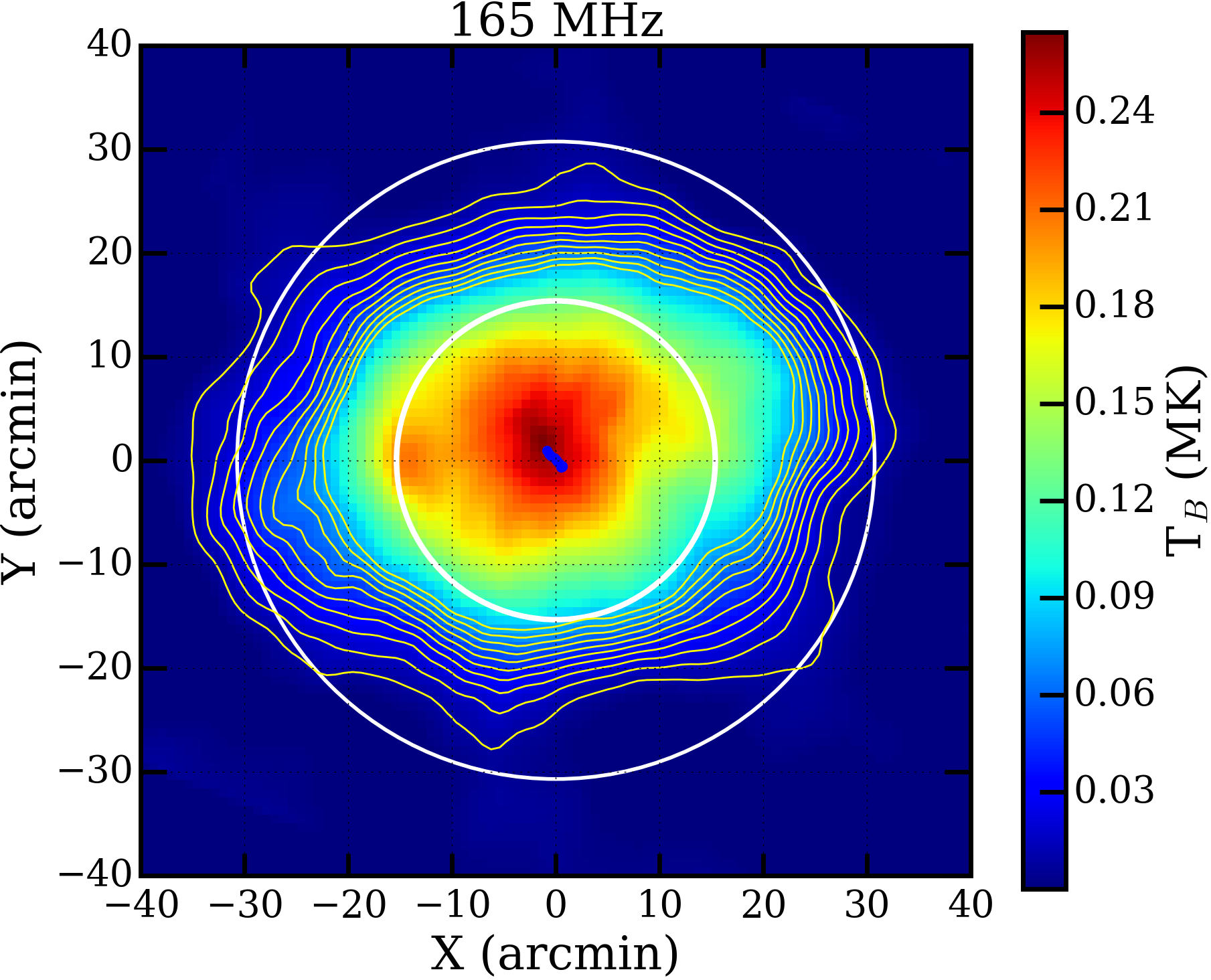}
 \includegraphics[trim = 0cm 0cm 0cm 0cm,width=0.30\textwidth]{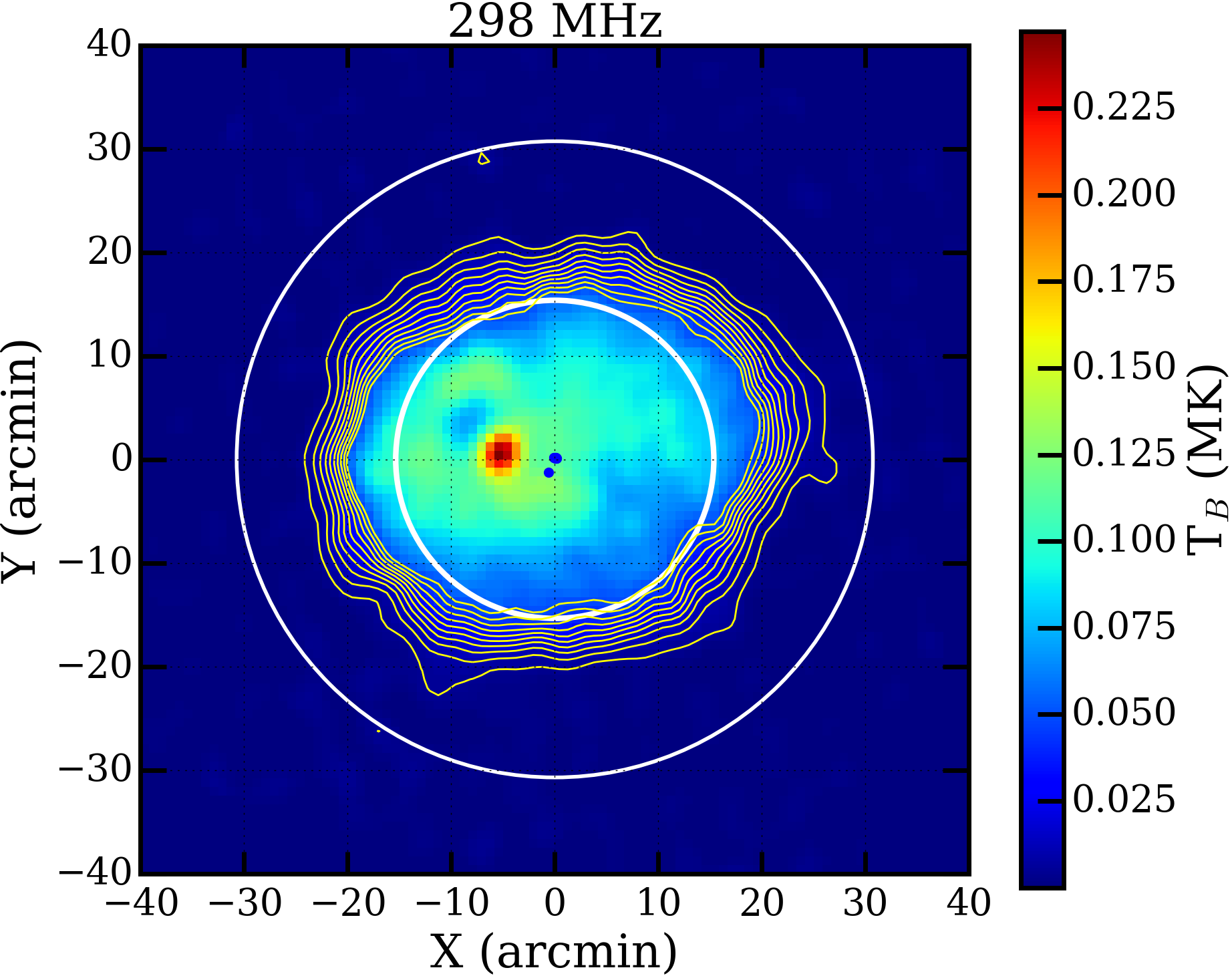}
 \caption{Solar disk images at 101, 165, and 298~MHz represent thermal emission at 03:48~UT. The yellow contours show thermal emission plotted as 0.4-4.0\% of the peak flux at that frequency. The two white circles shown in each panel have radii 1 R$_\odot$ and 2~R$_\odot$, the smaller one represents the limb of the optical solar disk.} \label{fig5_contours_fit}
 \end{center}
 \end{figure*}

Figure~\ref{fig5_contours_fit} shows radio images at three frequencies, which represent thermal emission at the radio quiet time at 03:48~UT. From the radio observations, it is known that the solar disk emission in the frequency range from 101~MHz to 298~MHz is produced by free-free emission. 

The free-free emission disk is best captured at the lower contour levels. The center of the Sun at each frequency was computed by fitting ten contours (contours plotted as 0.4-4.0\% of the peak flux at that frequency), which are shown in yellow. For a point of reference, the two white circles shown in each panel have radii 1 R$_\odot$ and 2~R$_\odot$, the smaller one represents the limb of the optical solar disk. As we expect, the size of the radio solar disk increases with decreasing frequencies, that is, the radio solar disk at 101~MHz is bigger than the radio solar disk at 298~MHz. This further confirms that the emission at higher (lower) frequencies originated at lower (higher) heights in the solar atmosphere. The blue dots represent the computed centroids.


The root mean square (RMS) was computed for the centroid positions at each frequency on the two Cartesian ($X$ and $Y$) and the radial coordinates ($r$) in the units of radio pixel (1~pixel = 50$\arcsec$). Table~\ref{centroid_table} shows the standard deviation $\sigma$ of the position of the center of the optical solar disk in the Cartesian and radial directions. All RMS values are within one pixel, demonstrating the robustness in calculating the position of the center of the Sun. Furthermore, the center of the optical disk and center of the radio solar disk were identified and coaligned with respect to EUV images.


 \begin{table}[!hbtp]
  \centering
 \caption{Positional variation of centroids at quiet times.}
 \resizebox{9cm}{!} {
 \begin{tabular}{cc c c c c c}

 \hline
  
Frequency   &$\sigma_{X}$   & $\sigma_{Y}$   & $\sigma_{r}$ & B$_{maj}$& B$_{min}$\\

(MHz) &(pixels) &(pixels) &(pixels)  & (arcsec)&(arcsec) \\


\hline


101               &0.9         &0.15        &0.91  & 312 & 246 \\

165               &0.59        &0.63        &0.86  & 240 & 204\\

298               &0.25        &0.51        &0.57  & 210 & 198\\

\hline

  \end{tabular} \label{centroid_table}
}
\tablefoot{The positional variation of centroids at quiet times is shown in Fig.~\ref{fig5_contours_fit} where 1~pixel = 50$\arcsec$. It is important to note the following: $\sigma_{r}^{2} = \sigma_{X}^{2} + \sigma_{Y}^{2}$.  The dimensions of half power beam width of point spread function (PSF) are shown. Area of the PSF = $\pi$B$_{maj}$B$_{min}$, where B$_{maj}$ and B$_{min}$ are major and minor axes, respectively.}
\end{table}

\subsection{Spatial-temporal study}
\label{radio_jet_compare}

\begin{figure}[!hbtp]
 \begin{center}
 \includegraphics[trim = 0cm 0cm 0cm 0cm, width=0.45\textwidth]{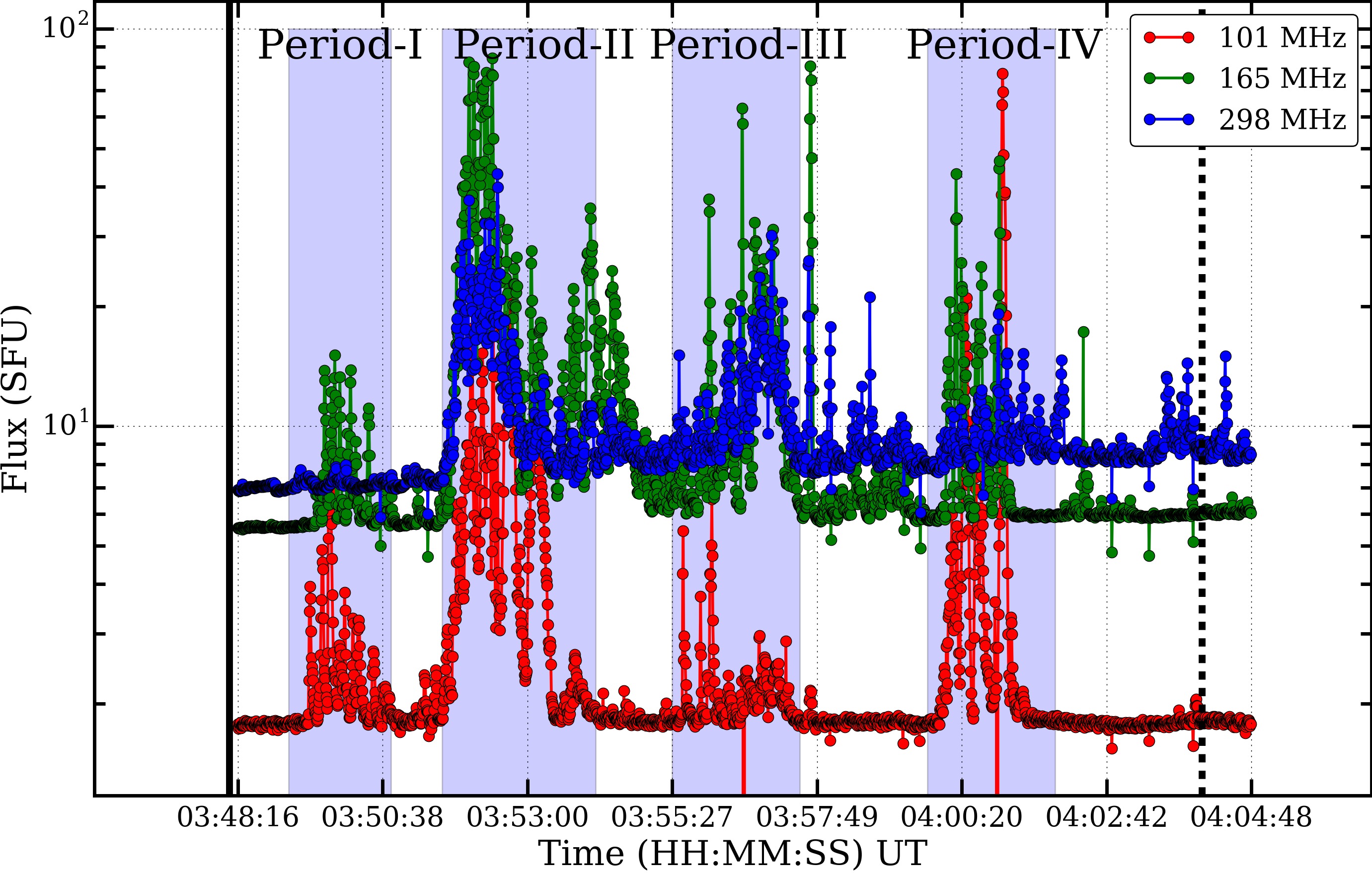} 
 \includegraphics[trim = 0cm 0cm 0cm -1.0cm, width=0.4\textwidth]{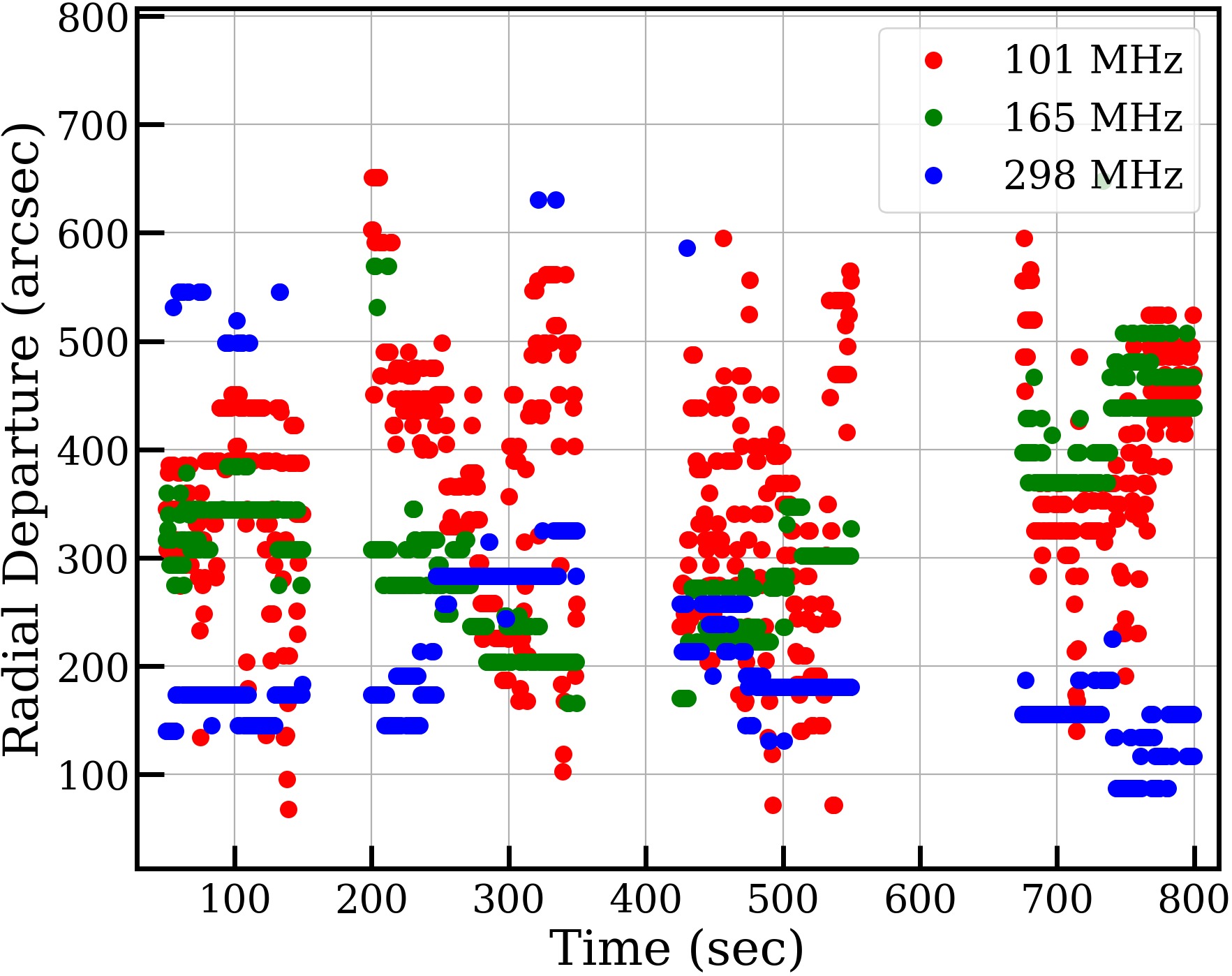}  
 \includegraphics[trim = 0cm 0cm 0cm -1.0cm, width=0.4\textwidth]{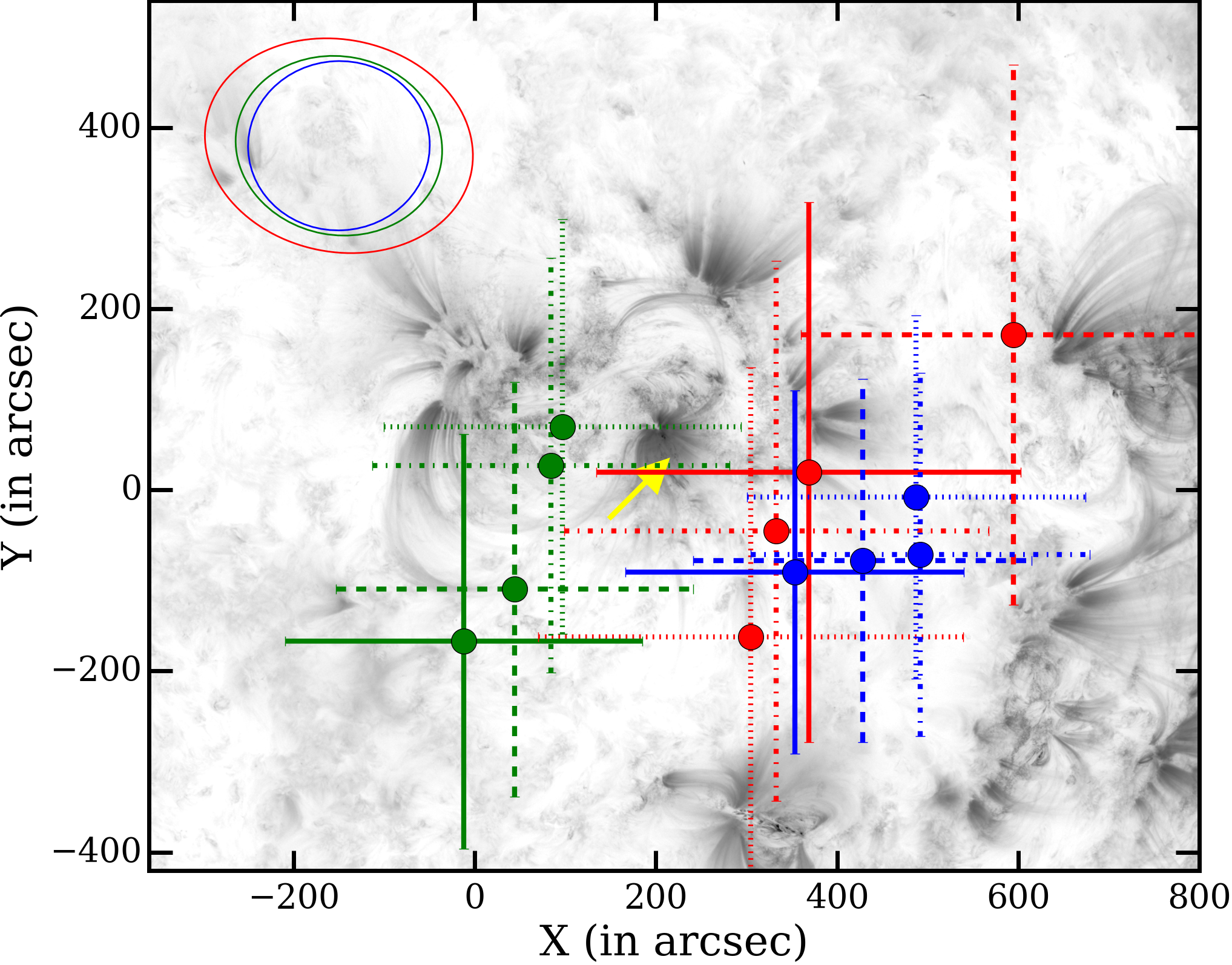}
\caption{Top panel: Temporal evolution of radio fluxes obtained from MWA spectrum for three selected bands. The solid and dashed vertical black lines indicate the start and end times of the jet event respectively. The purple shaded vertical bands indicate the periods of the four dominant radio bursts. Middle panel: Distance (in arcsecs) between location of jet and centroid of radio source for each of four bursts. Bottom panel: Location of radio sources with respect to jet position for different periods and frequencies. The blue, green, and red colors correspond to 298, 165, and 101~MHz respectively. The line styles solid (-), dashed (--), dashed-dotted (-.) and dotted (..) represent periods I-IV, respectively. The ellipses shown on the top-left are the beam sizes at 101 (red), 165 (green), and 298 (blue) MHz.} \label{fig6_burst_analysis}
 \end{center}
 \end{figure}

During the jet evolution, it was observed that the signatures of type III radio bursts appeared in columns in the MWA spectrum. Figure~\ref{fig6_burst_analysis} (top panel) shows the frequency averaged light curves for 101, 165, and 298~MHz. The spiky nature of the emission indicates nonthermal emission. Based on the high prevalence of impulsive nonthermal emission, we classified the four periods of bursts as I to IV (shown as purple patches in Fig.~\ref{fig6_burst_analysis}. top panel). Details about these periods and their duration are given in Table~\ref{Tab:band_table}. In addition to the burst periods, we observed impulsive emission outside of these bands. However, the majority of the emission comes in periods I-IV.

In Fig.~\ref{fig6_burst_analysis}, the black lines represent the start (solid line) at 03:48~UT and end (dashed line) times at 04:03~UT of the EUV jet. The start time of the jet is close to the time where period~I starts, whereas the end time of the jet continues after period~IV. The period of the evolution of the jet covers the duration from start of period~I to the end of period~IV. The onset of type III radio burst starts at 03:49:28~UT, 100~sec after the EUV eruption onset
(03:48~UT). The strength of impulsive emission are strongest in the 167-270~MHz band and significantly weaker outside this band.



\begin{table}[!hbtp]
\centering
\caption{Classification of dominant radio bursts.}
\resizebox{6cm}{!} {
\begin{tabular}{lccc}

\hline
 
 
period  &Duration       &Start time     & End time \\
        &(sec)          &(UT)           &(UT)  \\


\hline


I       &100            &03:49:06       &03:50:46  \\
II      &150            &03:51:06       &03:54:31  \\
III     &125            &03:55:21       &03:57:26  \\
IV      &275            &03:59:31       &04:04:06  \\

\hline

  \end{tabular} \label{Tab:band_table}
}
\end{table}


\begin{figure*}[!hbtp]
\begin{center}
\includegraphics[trim = 0cm 0cm 0cm -1.0cm, width=0.4\textwidth]{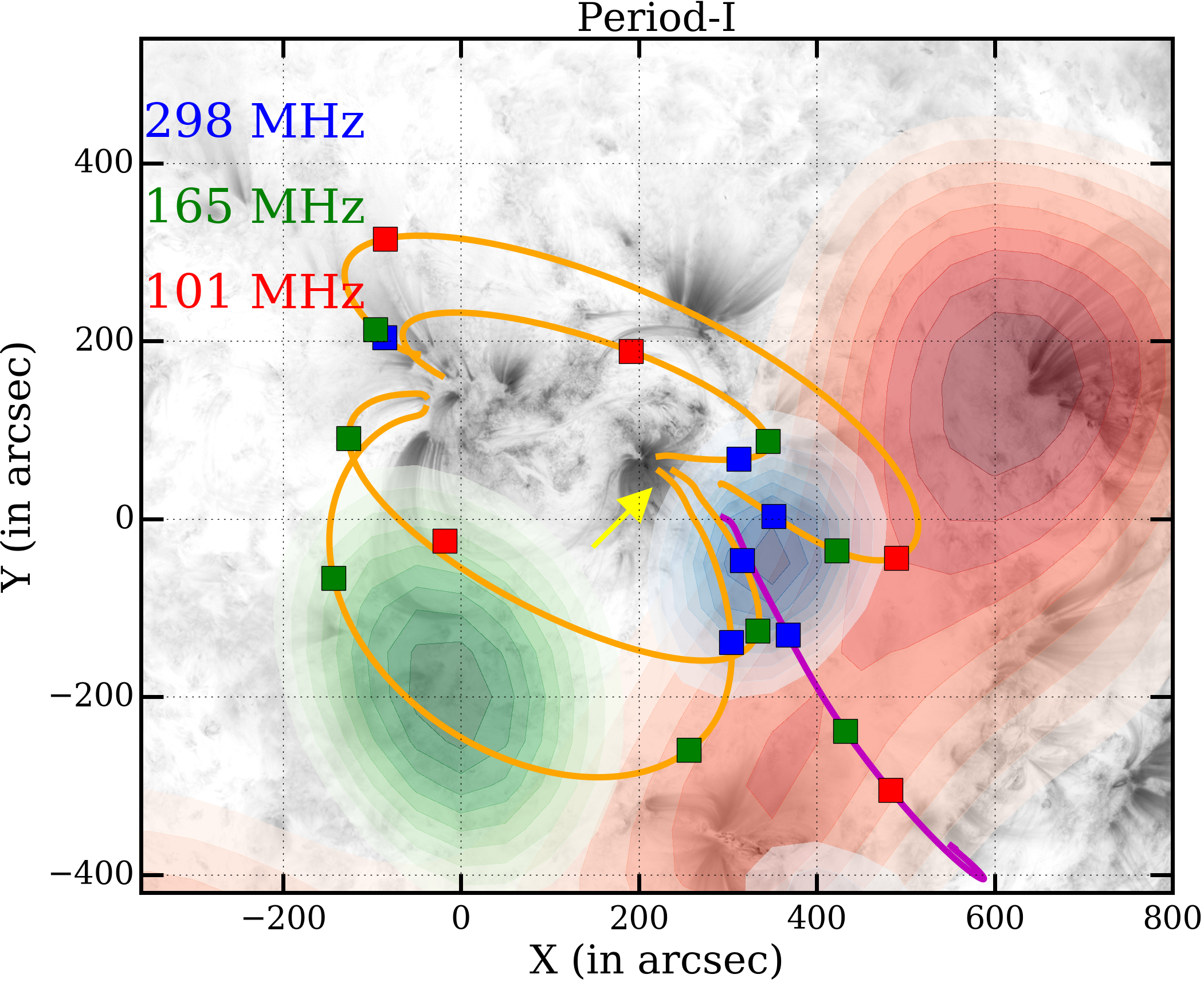} 
\includegraphics[trim = 0cm 0cm 0cm -1.0cm, width=0.4\textwidth]{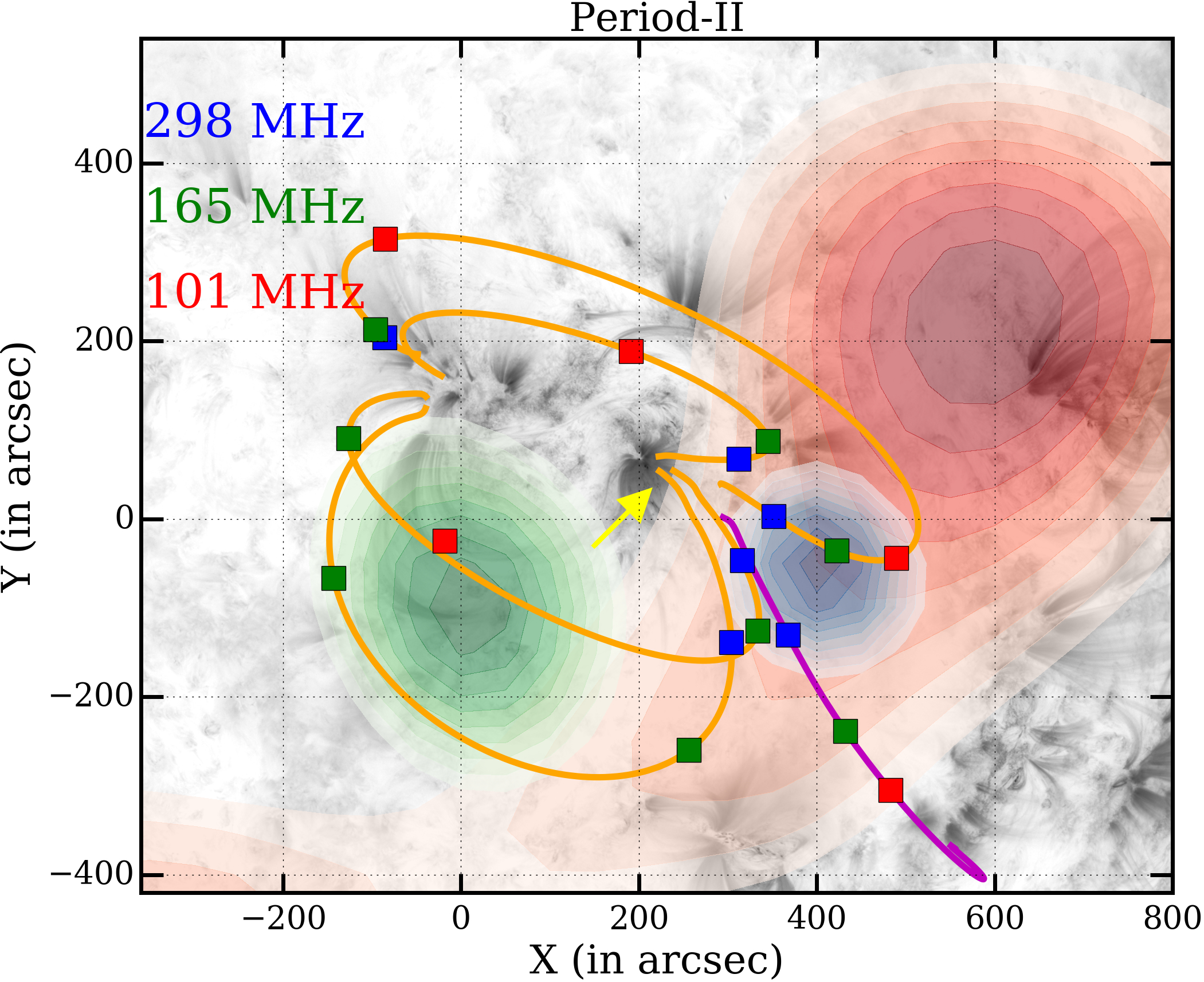}
\includegraphics[trim = 0cm 0cm 0cm -1.0cm, width=0.4\textwidth]{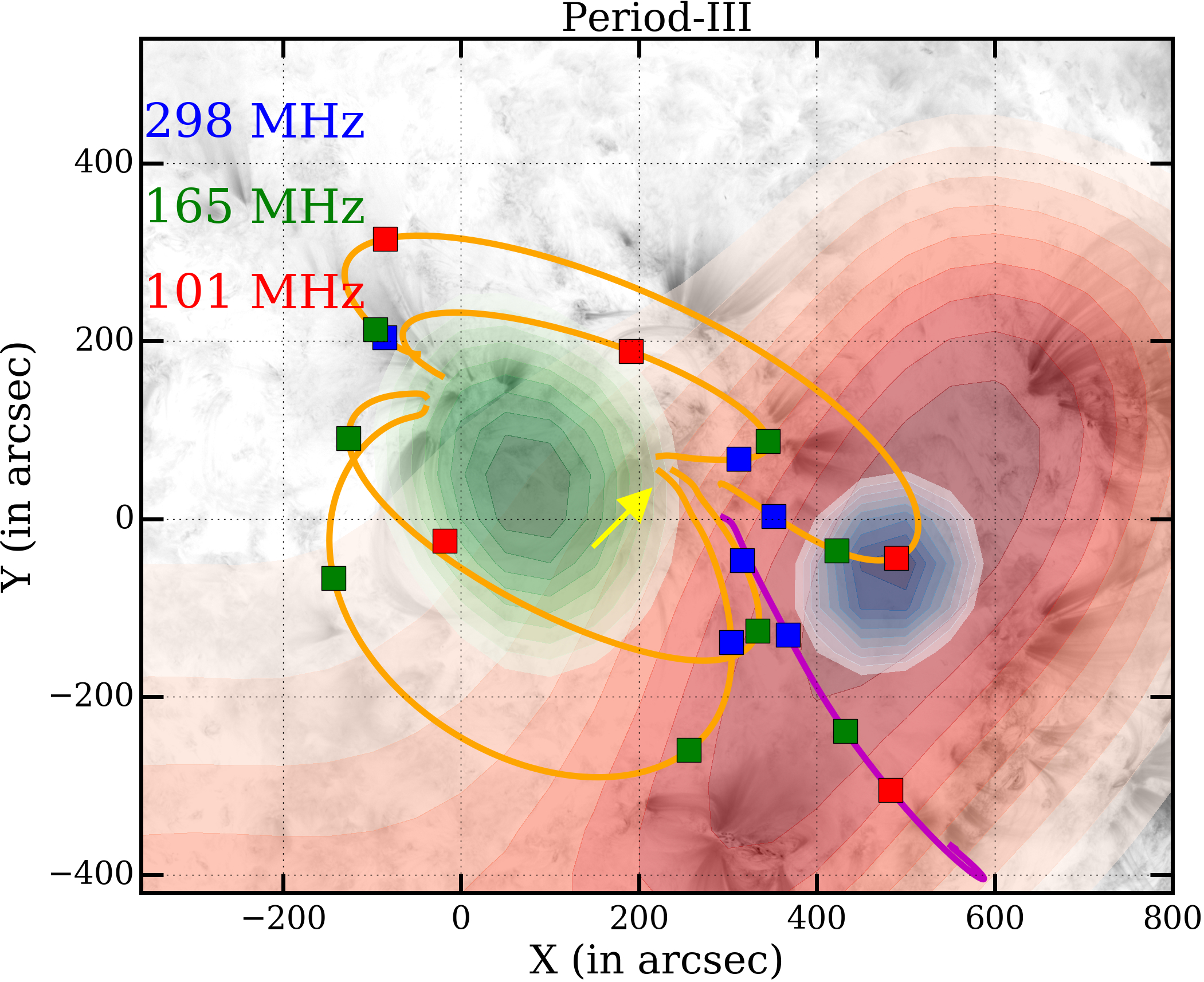} 
\includegraphics[trim = 0cm 0cm 0cm -1.0cm, width=0.4\textwidth]{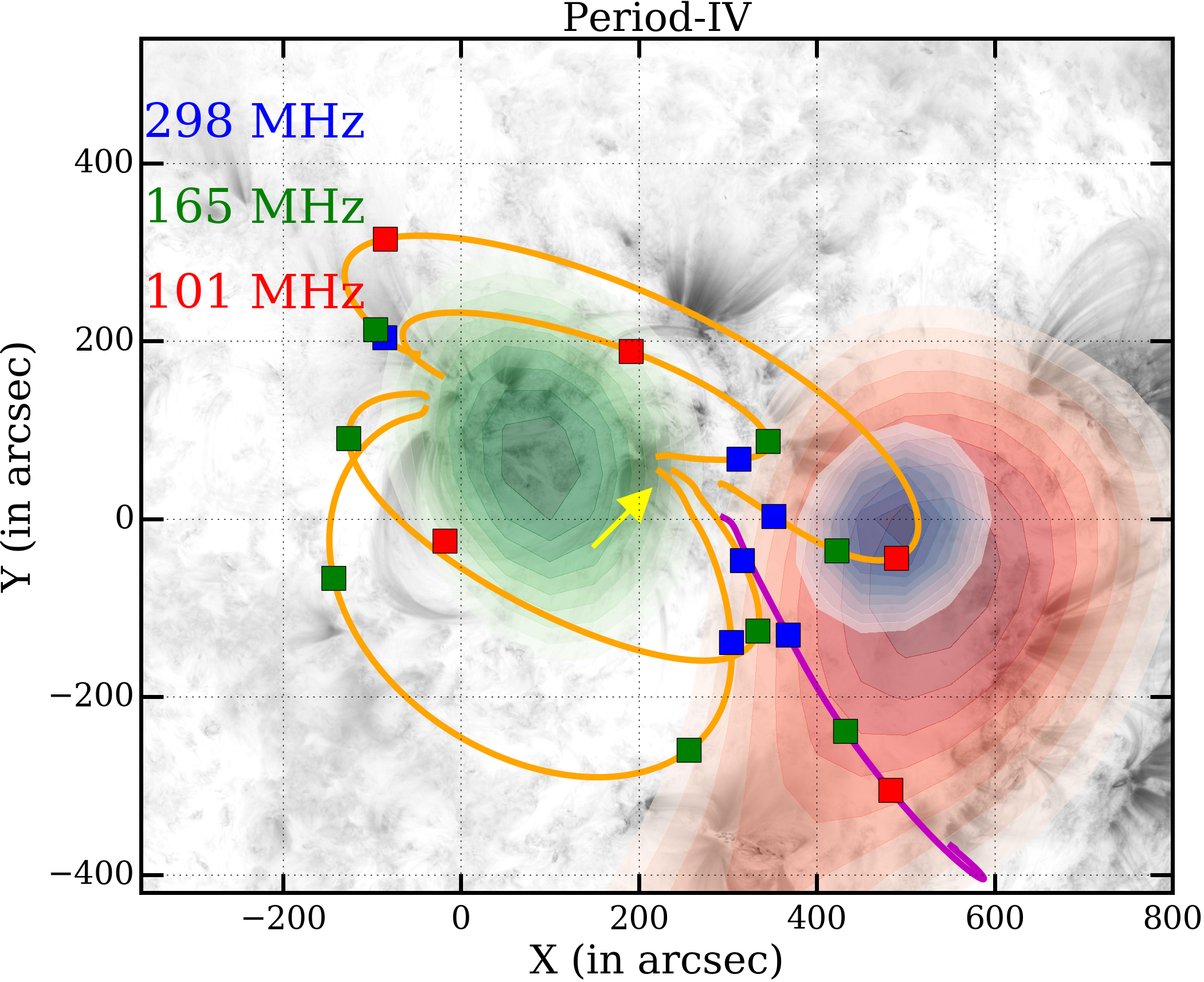} 
\caption{AIA 171~{\AA} image at 03:51:47~UT and red, green, and blue overplotted contours (50, 55, 60, 65, 70, 75, 80, 85, 90, and 95\% of peak of flux) represent 101, 165, and 298~MHz radio frequencies, respectively. The yellow arrow indicates the location of the jet eruption. Orange and magenta lines indicate closed and open field lines, respectively, which are derived from the extrapolation. The red, green, and blue colored squares show the locations along extrapolated magnetic loops where the local fundamental plasma frequency is expected to be 101, 165, and 298~MHz (see text), respectively.} \label{fig7_radio_sources_imag}
\end{center}
\end{figure*}

The radio images at 101, 165, and 298~MHz show the presence of bright compact radio sources in the vicinity of the jet eruption. Figure~\ref{fig7_radio_sources_imag} shows the period averaged radio sources for three frequency bands overplotted on the AIA 171~{\AA} images during the jet eruption. The centroids showed a significant positional variation for the four periods of bursts. For period I, it is observed that the radio sources at 298~MHz were closer to the jet eruption site, whereas radio sources at 165~MHz and 101~MHz were spread over a larger area and appeared farther from the jet eruption site. For subsequent periods II-IV, the 165~MHz source roughly moves along the direction of the yellow arrow, whereas the 101~MHz source is much more inclined as compared to the PSF and shows more complicated variations.



Table~\ref{Tab:centroid_table} shows the mean and RMS of the radial distance of the dominant radio sources from the EUV eruption site. Figure~\ref{fig6_burst_analysis} (bottom panel) shows the location of the centroids for the four period of bursts for all three frequencies from the distance of the jet eruption. The maximum variation of the centroids for the four period of bursts was obtained for 101~MHz at about $\sim$117". Larger positional fluctuations of 124" were seen for 298~MHz within Band-I (also seen in Fig.~\ref{fig6_burst_analysis}, bottom panel). However, it is important to note that these shifts are within the two to three PSF region (see Table \ref{centroid_table} and \ref{Tab:centroid_table}). 

In a type-III burst, the energetic electrons travel along the magnetic field lines, interacting with the local Langmuir waves and creating the radio sources along the field lines. The emission at a frequency $\omega$ comes from a coronal height where the local plasma frequency (or its harmonics) are equal to $\omega_p$. 
In a simplified scenario, the locations of the radio sources at various $\omega_p$ systematically align with the direction of propagation of the energetic electrons, primarily along the field lines \citep{Bin2015}. Some of these emission spots are shown on the extrapolated magnetic field lines (see Section~\ref{section5}) in Fig.~\ref{fig7_radio_sources_imag}.



The observed radio sources shown in Fig.~\ref{fig7_radio_sources_imag} do not follow a simplistic picture. Radio sources, particularly at low-frequencies are expected to show deviations from their true positions. Propagation effects like refraction, scattering, and wave-ducting could be responsible for these shifts. In the present event, due to it's central location, the refraction is expected to be minimal. Scattering is a major phenomenon in shifting the positions of radio sources \citep{Steinberg71}. Many studies done on scattering have revealed shifts due to density fluctuations (e.g. \citealt{Thejappa94, Krupar18}). The density inhomogeneities can occur due to the uneven distribution of plasma along the loop. They impact the apparent radio source location, mostly when present in the vicinity of the emission location. These shifts (see Table~\ref{Tab:centroid_table}) increase with decreasing frequencies \citep{Robinson83}. The scenario becomes more complex in an eruption due to dynamic changes associated with the emitting source itself. The emitting source moves, thus creating a scattered traveling apparent radio source. Figure~\ref{fig6_burst_analysis} (bottom panel) shows the source movement in the imaging plane. This direction of the motion from period I to IV is radially away for 298~MHz, while it becomes irregular for 165 MHz and 101~MHz. The density inhomogeneities can vary temporally from jet eruption at MHD timescales. However, the abrupt large shifts observed (298~MHz in Fig.~\ref{fig6_burst_analysis}) are too fast for a scattering process. Two clear cases of clustering in the 298 MHz sources, one at the radial distance 100$\arcsec$ to 300$\arcsec$ and another at 500$\arcsec$ to 600$\arcsec$, suggest a secondary source. Eruption processes are known to be fragmentary and can be produced in the secondary sources away from the eruption site. Furthermore, there are fine temporal shifts within each period, which are captured in the smeared extended radio sources (Fig.~\ref{fig6_burst_analysis}, especially for 101~MHz). The smearing occurs due to the image averaging over the duration of the respective periods. The combination of scattering and intrinsic source movement are responsible for the complex motion.



 \begin{table}[!hbtp]
\centering
\caption{Positional variation of centroids at radio frequencies for
four periods.}
\resizebox{7cm}{!} {
\begin{tabular}{l c c c}
\hline 

Bands   &101 MHz   & 165 MHz   & 298 MHz \\
        &(arcsecs) &(arcsecs) &(arcsecs)  \\

\hline
&&&\\ 

I             & 360$\pm$78        & 332$\pm$24 & 212$\pm$124  \\
II            & 405$\pm$117       & 256$\pm$61 & 254$\pm$65  \\
III           & 330$\pm$112       & 267$\pm$44 & 203$\pm$40 \\
IV            & 385$\pm$83        & 420$\pm$46 & 142$\pm$31  \\
&&&\\ 
\hline

Coronal Heights & 0.50 R$_{\odot}$ & 0.32 R$_{\odot}$&0.13 R$_{\odot}$\\

\hline
\end{tabular} \label{Tab:centroid_table}}
\tablefoot{The positional variation of centroids at radio frequencies for the four periods is shown in Fig.~\ref{fig7_radio_sources_imag}. The last row shows the coronal heights corresponding to the three frequencies based on \citet{Newkirk61} electron density model, assuming 4$\times N_{\rm e}$ active region coronal densities. It is important to note that the coronal heights are calculated from the solar surface.}
\end{table}

\subsection{Temporal evolution of radio brightness temperature}  
\label{brightness_temperature}

\begin{figure*}[!hbtp]
\begin{center}
\includegraphics[trim = 0cm 0cm 0cm 0cm,width=0.9\textwidth]{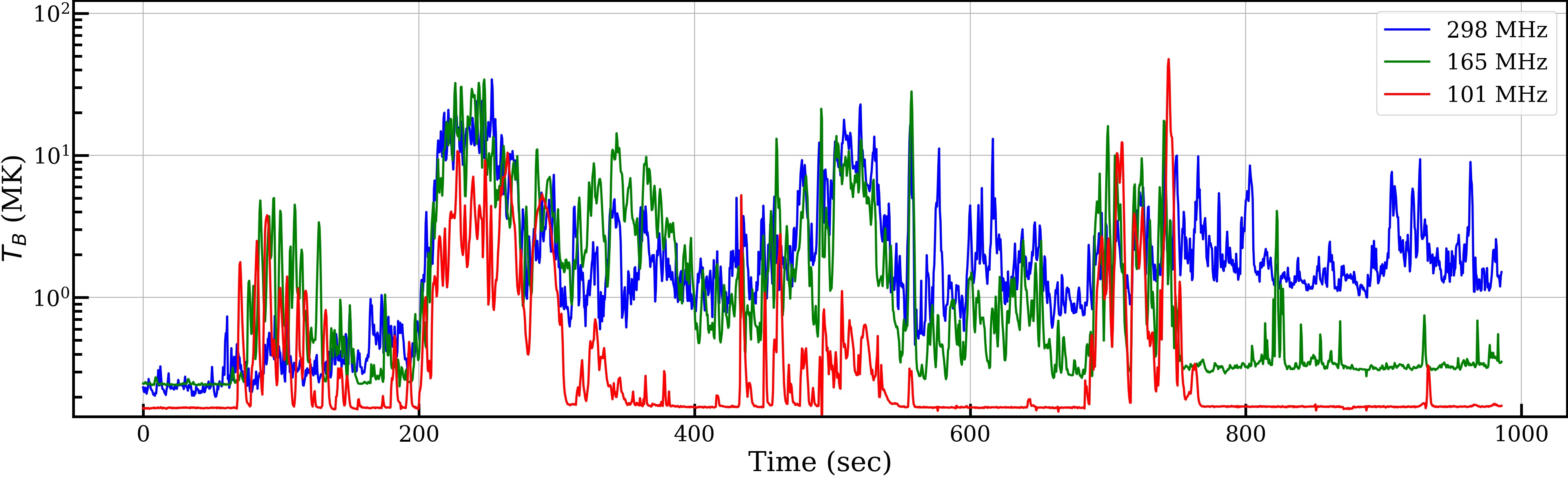} 
\caption{Temporal evolution of brightness temperature $T_B$ derived from dominant radio sources, computed over 90\% contour levels of radio images.} \label{fig8_meanTb}
\end{center}
\end{figure*}

Figure~\ref{fig8_meanTb} shows a temporal evolution of the brightness temperature $T_{B}$ of the radio sources observed at 101, 165, and 298~MHz. The radio quiet time at 03:48~UT was chosen as the start time for this plot. Before the nonthermal type~{\rm III} burst, the $T_{B}$ for the three frequencies was quite similar and close to the quiet Sun thermal temperature \citep{Oberoi17}. The brightness temperatures were computed over 90\% contours. The impulsive emission seen in the time series represent the nonthermal emission at different frequencies. The
time series shows a higher prevalence of nonthermal emission at the 298~MHz band, which decreases with decreasing frequency (higher height). The brightness temperature light curve shows the presence of numerous short lived impulsive structures, suggesting the complexities in the energies emitted from the plasma emission mechanism.

\section{Modeling the local and global magnetic field} 
\label{extrapolation}

The jet activity was examined using images in the AIA 211~{\AA}
channel. The area of origin for the eruptions was just south of a
compact sunspot, covering approximately an area of $\sim$40 arcsec$^{2}$ from the footpoint of the jet. The brightening was observed at the footpoint, which was associated with the southern end of the light bridge region. A small coronal hole was located to the southeast of the sunspot.

In this section, we investigate the magnetic structures associated with the source region of the jet and their connectivity with the open magnetic field structure and the nonthermal type~{\rm III} radio bursts. The aim was to check if there was a possible magnetic connection between the relatively small structure where the brightening occurred and the very large scale of the open magnetic field region.


The coronal magnetic field can be reconstructed from the remote observations of the photospheric magnetic field. The corona is force-free on temporal and spatial scales. The force-free assumption is also true for plasma in thermal equilibrium, which is dominated by the magnetic field. Additionally, under this assumption, the Lorentz force resulting from the interaction between field and currents vanishes everywhere. The force-free extrapolation technique solves the corresponding set of equations under the constraint that the field reproduces the observed photospheric values. 

\begin{figure}[!hbtp]
\begin{center}
\includegraphics[trim = 0.1cm -1.0cm 0.1cm -1.0cm,width=0.43\textwidth]{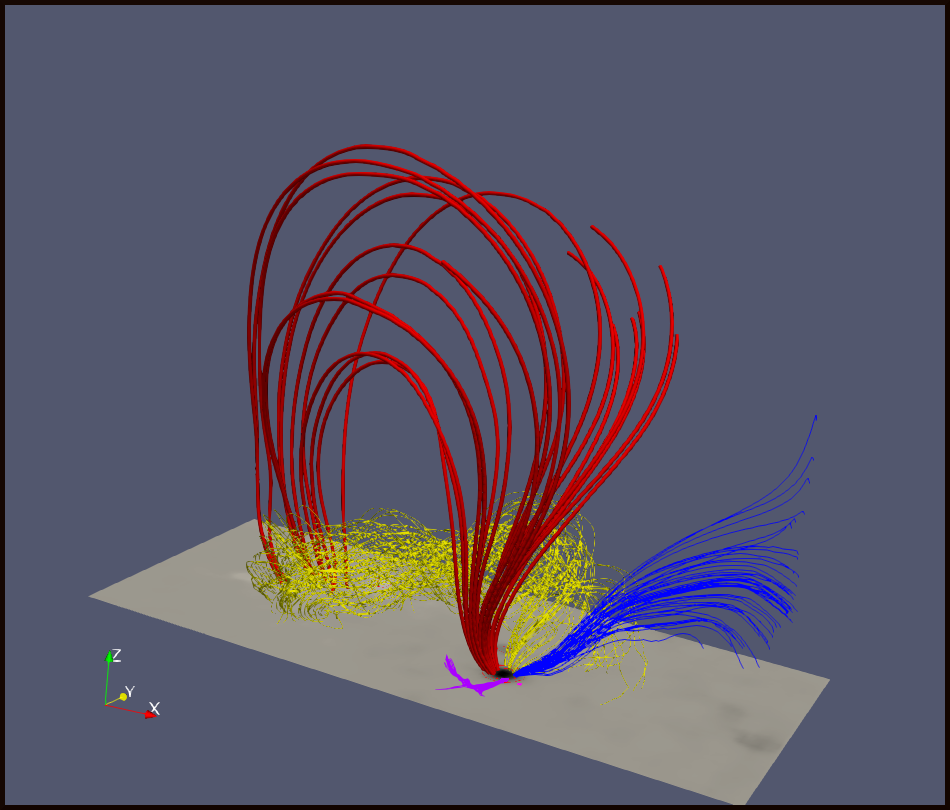}
\caption{Global configuration of active region was obtained by using a NLFFF extrapolation. The red field lines represent a wide envelope of the active region, yellow lines represent flux rope structure, blue lines represent the open field lines, and pink field lines represent the source region of the jet. The active region patch of the vector magnetic field from the HMI SHARP series is shown in a gray color and the black dot represents the negative polarity sunspot.}
\label{fig10_ar_mag_field}
\end{center}
\end{figure}

Because of the numerical efforts required for extrapolation, a multi-scale approach is considered here. This involves two different types of magnetic extrapolations: NLFFF and PFSS extrapolation. Firstly, the local configuration for the magnetic field of the jet was obtained using the general nonlinear force-free extrapolation technique. The method was applied to a field of view covering the entire active region area. The high-resolution (about 0.5$\arcsec$) photospheric vector magnetic field data was obtained from the Space weather HMI Active Region Patch (SHARP)\footnote{http://jsoc.stanford.edu/doc/data/hmi/sharp/sharp.htm} series of the Helioseismic and Magnetic Imager (HMI) \citep{Scherrer12, Schou12} on September~3, 2013 at 03:36~UT, which covered the entire active region and nearby area. 

The NLFFF magnetic extrapolation was obtained by using the method described in \cite{Valori10} and references therein. The analysis was performed in a similar way to that reported in section~6.1 of \cite{Polito17}. Figure~\ref{fig10_ar_mag_field} shows the NLFFF model of the lower coronal magnetic field structure associated with the active region and the source region of the jet. 

\begin{figure}[!hbtp]
\begin{center}
\includegraphics[trim = 0.1cm 0.1cm 0.1cm -1.0cm,width=0.43\textwidth]{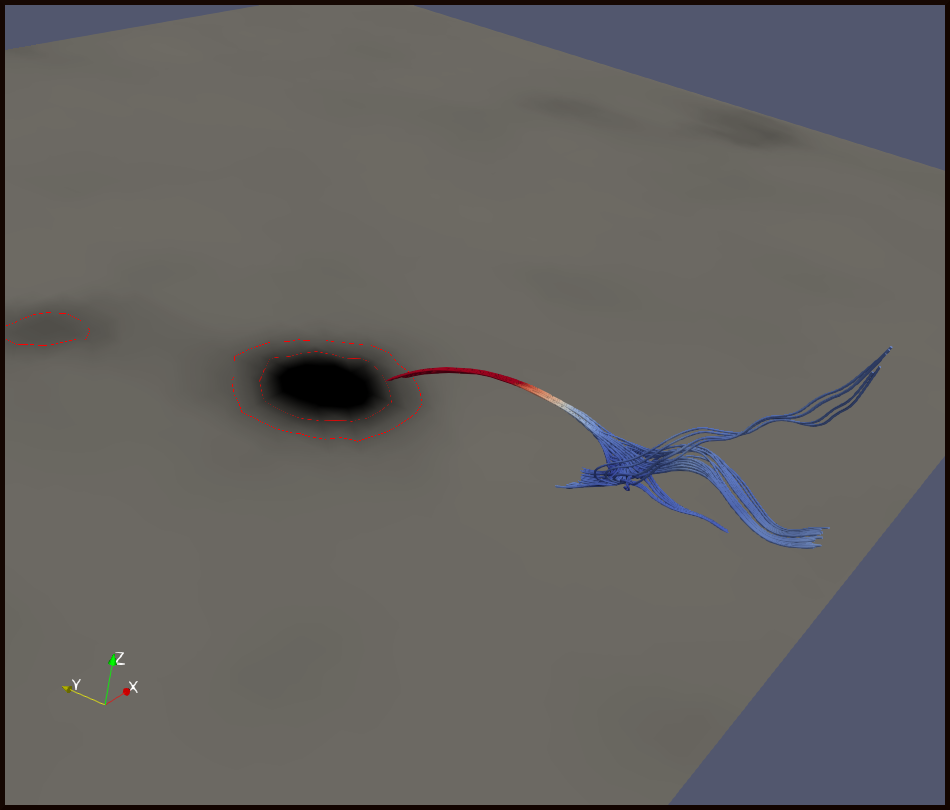}
\includegraphics[trim = 0.1cm -1.0cm 0.1cm -1.0cm,width=0.43\textwidth]{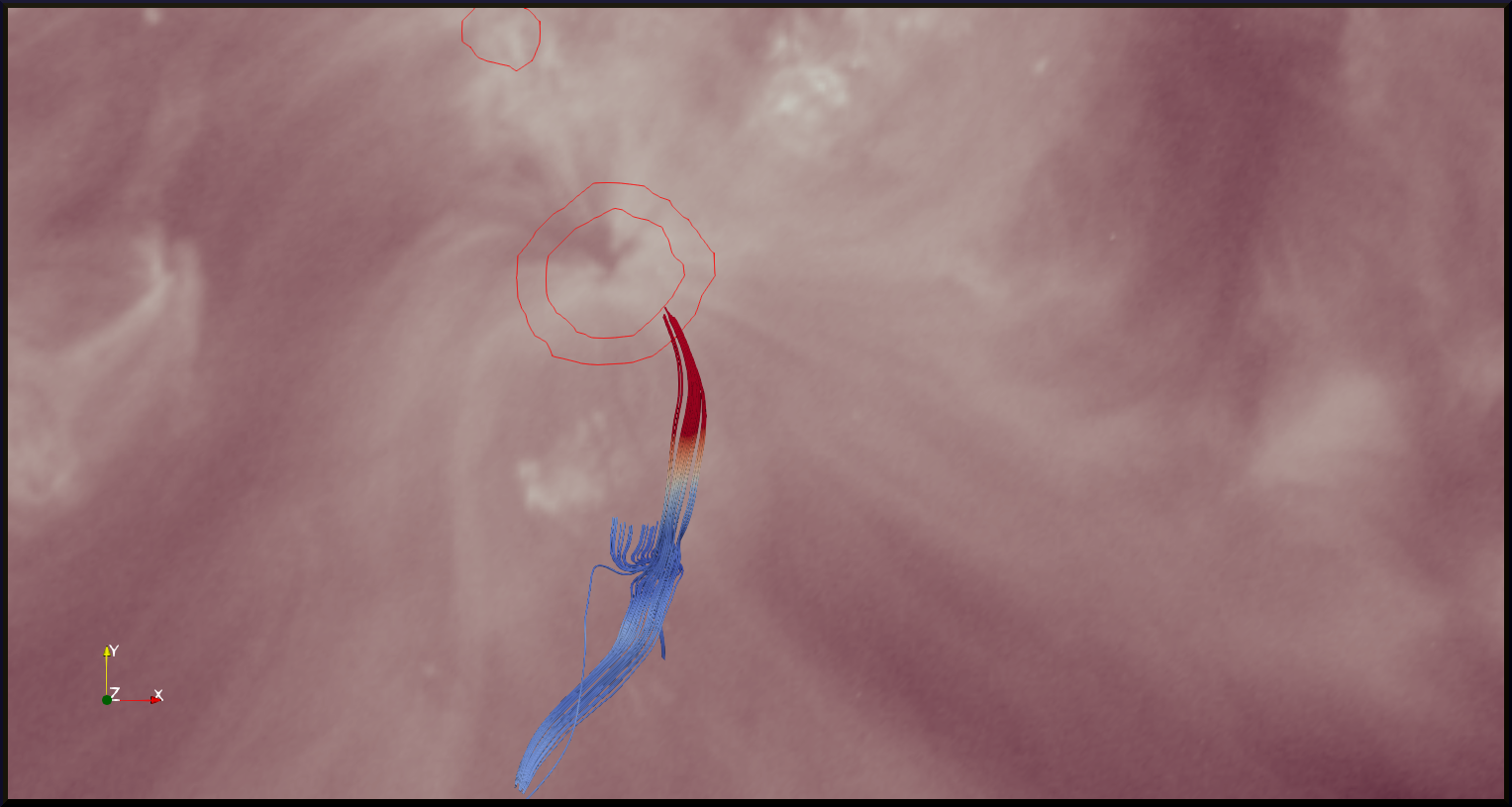}
\caption{Top panel: Zoomed view of null point field line structure associated with source region of jet. Bottom panel: Similar magnetic topology is overplotted on AIA 211~{\AA} image. The vertical magnetic field isolines of 200 G and 500~G are shown as red contours. The red and blue colors represent the magnetic field 0 $<$ $|$B$|$ $<$ 200~G.} \label{fig11_jet_fig13}
\end{center}
\end{figure}


\begin{figure}[!hbtp]
\begin{center}
\includegraphics[trim = 0.1cm 0.1cm 0.1cm 0.1cm,width=0.43\textwidth]{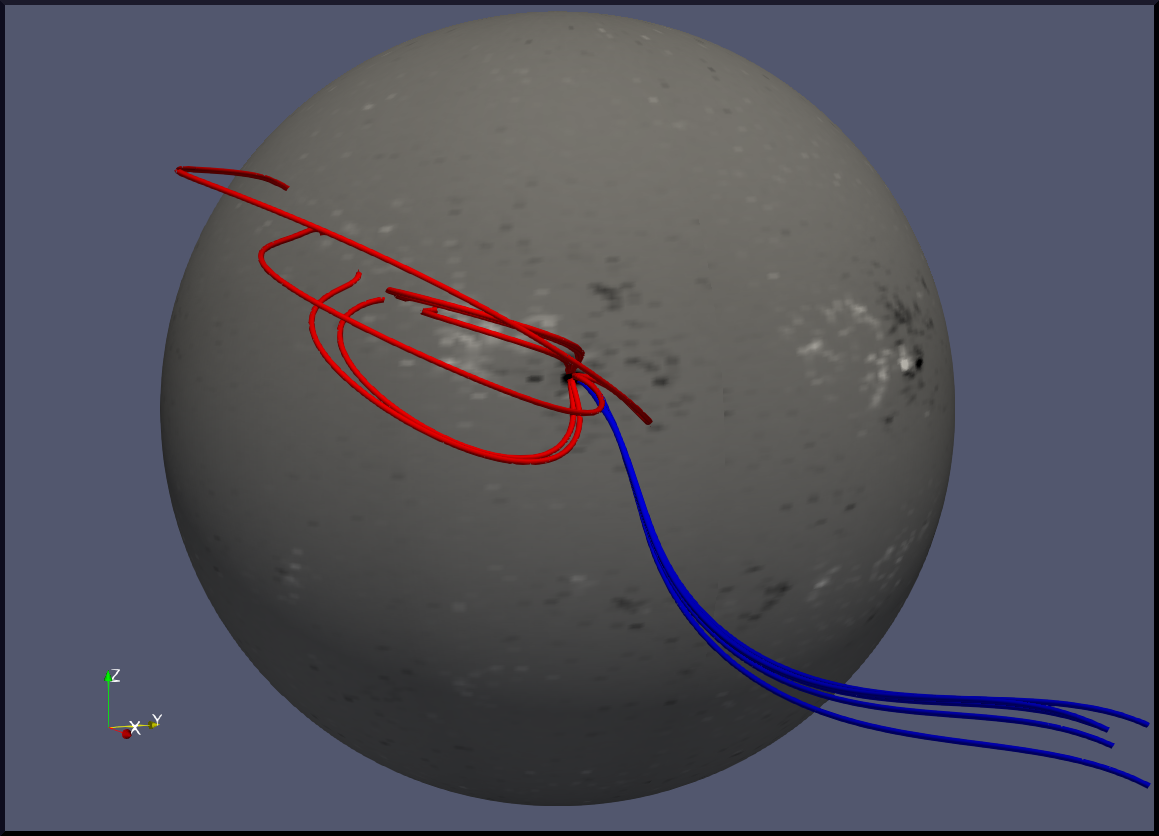}
\includegraphics[trim = 0.1cm -1.0cm 0cm -1.0cm,width=0.45\textwidth]{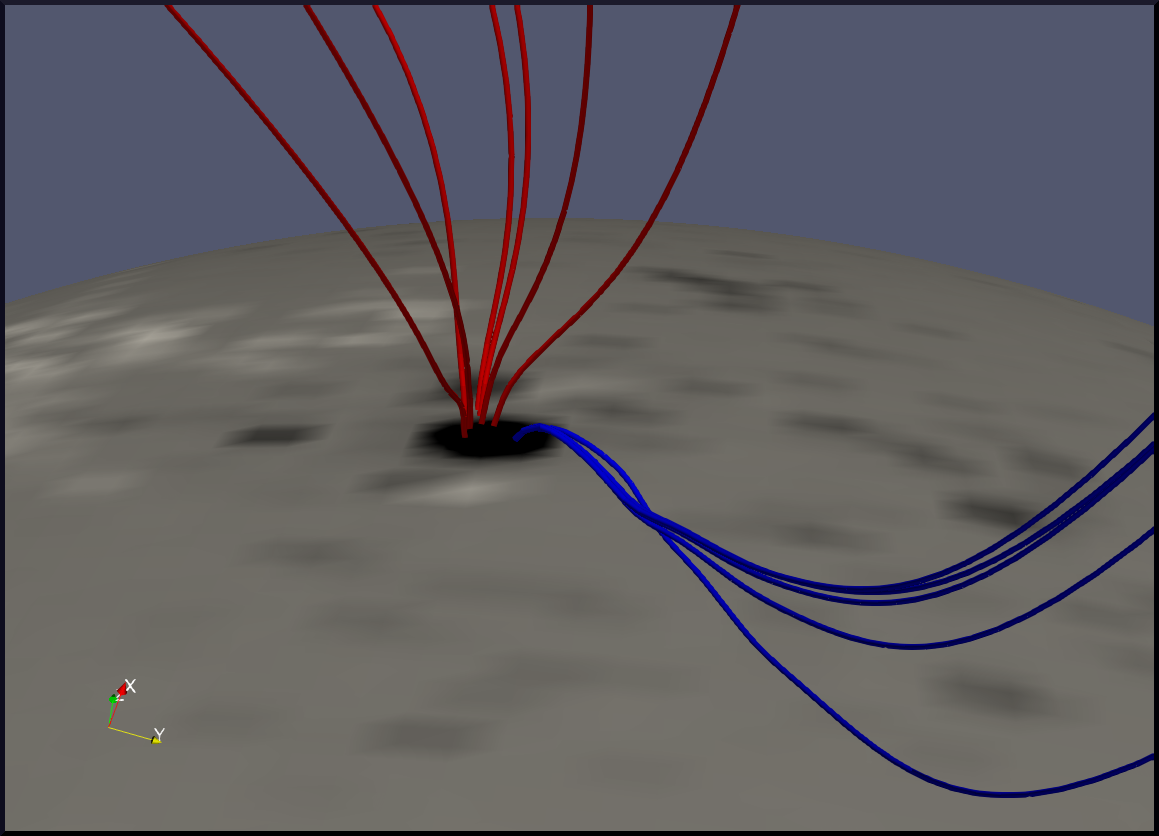}
\caption{Top panel: Selected field lines from PFSS extrapolation. Bottom panel: Zoomed view of active region field representing location of footpoint of selected field lines on sunspot.} \label{fig12_jet_fig14}
\end{center}
\end{figure}

A zoomed view of the source region of the jet is shown in Fig.~\ref{fig11_jet_fig13} (top panel). The external spine of the null was rooted in the outer penumbra of the sunspot, which was in the proximity of the light bridge. This is the same region where the footpoint of the jet is rooted (which is shown in Fig.~\ref{fig1_aia_context_img} with a cyan line). 



The spine surmounted a convection cell where flux was seen continuing to emerge. The fan field lines are also shown where some deformation was present, which was likely due to the proximity of the vertical boundary of the extrapolation domain. The bottom panel shows the AIA 211~{\AA} image and the magnetic structure associated with the jet is overplotted. The isolines of 200 \ G and 500~G of the vertical magnetic field are overplotted as big and small red circles, respectively, on the negative polarity sunspot. A good qualitative agreement between the rooted field lines of the null and the location of the brightening was obtained.

From the sunspot in the proximity of the spine footpoint, field lines are present (shown as blue lines in Fig.~\ref{fig10_ar_mag_field}) that have a very different orientation and leave the extrapolation box from the lateral side. These were possibly open field lines, but their nature could not be affirmed because of the small size of the extrapolation box. In order to test such a hypothesis, a large-scale PFSS extrapolation was carried out. 


To model the coronal magnetic field at a global scale, a PFSS
extrapolation was also computed. Such an approximation is suitable to
represent the average solar force-free magnetic field at large scales,
but it fails at modeling structures that are strongly influenced by the
presence of coronal currents. The radial magnetic field at the coronal
base required by the PFSS method was fixed based on the synoptic
magnetogram provided by the HMI instrument for CR-2141, resampled at
half resolution. The Finite Difference Iterative Solver (FDIPS),
developed by \cite{Toth11}, was run on a spherical grid with a source surface radius set at $2.5\,\rm{R}_\odot$, with 75 uniform radial bins, 1440 latitude bins (uniform in cosine of latitude), and 3600 uniform longitude bins.

Figure~\ref{fig12_jet_fig14} shows selected field lines from the PFSS extrapolation, traced from the same set of starting points of the lines selected for the NLFFF model in Fig.~\ref{fig10_ar_mag_field}. Here, red and blue indicate closed and open field lines, respectively. The figures show reasonable consistency between the topology of both models in terms of closed and open structures.

As in the NLFFF extrapolation, the red field lines represent an envelope field above the active region and they started from the eastern side of the sunspot (see Fig.~\ref{fig12_jet_fig14}). The blue field lines started from the western side of the spot, and are indeed open, in the sense that instead of connecting back to the photospheric surface, they intercept the upper boundary of the model, placed at two solar radii. 


\section{Comparison of the extrapolated magnetic field with radio sources}
\label{section5}


As previously mentioned, the radio sources ideally originate from the $\nu=\nu_p$ surfaces under simple stratified solar corona and under the absence of propagation effects. 
But as seen in the radio images, these assumptions are quite far from reality as meterwaves suffer from strong refraction and scattering in the corona. Figure~\ref{fig7_radio_sources_imag} shows the expected
position of the radio sources at 101, 165, and 298~MHz on the magnetic
field lines assuming the Newkirk coronal density model \citep{Newkirk61} (assuming 4$\times N_{\rm e}$ active region coronal densities). Using the magnetic field lines derived from extrapolation in Section~\ref{extrapolation}, a few large scale closed and open field lines were chosen for the representation. The computed source positions on the field lines show the diverse spatial locations adding to the complexity of the radio source location.


We observed the expected locations of the 298~MHz source close within a 1-2 PSF element in relation to its apparent observed locations. While the observed 100~MHz radio sources are located far from its expected location on the open magnetic field. This is expected as the longer wavelengths would be relatively more impacted by scattering and refraction \citep{Li12, Thejappa12}. However the 165~MHz source appears toward the east of the eruption site. Interestingly, the observed sources for 165~MHz appear to come from the near loop top regions of large scale closed magnetic fields. We know that wave ducting can also cause shifts in the radio source position under specific conditions of the emission angle with the coronal loop boundaries and make them appear near the apex of the loops \citep{Duncan79}. The radio waves can be ducted away from the original source location, later showing up near the apex of the loop (up to 500$\arcsec$ far) \citep{Bisoi18}.


The model regarding the extrapolation of magnetic fields is simplistic as it does not include the local small scale variations.
These local variations in magnetic fields along with density inhomogeneities can impact the apparent position of the radio sources. Overall, when compared with the observed MWA radio sources, the extrapolated magnetic field presents a contrast in the calculated radio source locations.  However, the observed sources are compatible with the observed contrast in extrapolated magnetic fields.

\section{Summary and discussion}  \label{discussion}


In this paper, we carried out a comprehensive investigation of an active region jet and a nonthermal type III radio burst by combining data from various observatories, such as EUV imaging data from AIA/SDO, radio imaging and spectrum from MWA, and photospheric magnetic data from HMI/SDO. The EUV jet showed a complex evolution of the thermal plasma, which is trapped in the complex magnetic structure of the active region. The changes in the topology of the coronal magnetic field were evident from the untwisting motion of the jet plasma along its axis. The plane-of-sky velocity of the AR jet was found to be $\sim$136~km/s. The EM analysis confirms that the hot plasma of temperature 5-8 MK was present at the footpoint region and also showed the presence of Fe~{\sc xviii} 93.93~{\AA} emission, whereas the spire was at a low temperature of 2~MK. The spire temperature was found to be in good agreement with the temperatures obtained in other studies of AR jets \citep{Mulay16, Mulay17a}. The electron number densities in the spire and footpoint were found to be an order of magnitude lower compared to the densities measured for other jets studied by \cite{Mulay16, Mulay17a, Mulay18}. However, these are lower limits.

A temporal and spatial correlation between the AR jet and nonthermal type III burst was confirmed using the radio imaging. The radio emission of the event showed a high prevalence of fine structures in T$_{B}$, frequency, and time, thus showing the complexities involved in the plasma emission mechanisms.

The type III burst mainly occur during four periods spanning the entire meterwave wavelength range corresponding to 0.13–0.54 R$_{\odot}$ (Table~\ref{Tab:centroid_table}). During these periods, radio emission took place over a wide range of frequencies, which can be interpreted as the source regions that are located at different coronal heights. The large differences (from 0.13-0.52 R$_{\odot}$ above the surface) between the emission heights of the three radio frequency bands suggest the presence of the open magnetic fields. These open magnetic field lines facilitate the movement of nonthermal electrons into the interplanetary medium and produce radio emission at lower frequencies (as seen in
Fig.~\ref{fig4_type3_spectrum}, middle and bottom panel). However, the above mentioned portrayal of the eruption phenomenon seems simplistic. Radio imaging shows significant shifts in the source location at different frequencies in no systematic manner. Such behavior strongly suggests the presence of propagation effects, such as refraction and scattering that are created by strong, dynamic density inhomogeneities.

The local configuration of the magnetic field at the location of the jet was studied using the NLFFF extrapolation. The analysis shows that the photospheric footpoints of the null point (see Fig.~\ref{fig11_jet_fig13}) were anchored at the location of the source brightening of the jet (see Fig.~\ref{fig1_aia_context_img}). The connectivity of the source region of the jet with open magnetic field structure was studied using the PFSS extrapolation.

\begin{figure}[!hbtp]
\begin{center}
\includegraphics[trim = 1.8cm 0.3cm 1.8cm 0.5cm, width=0.38\textwidth]{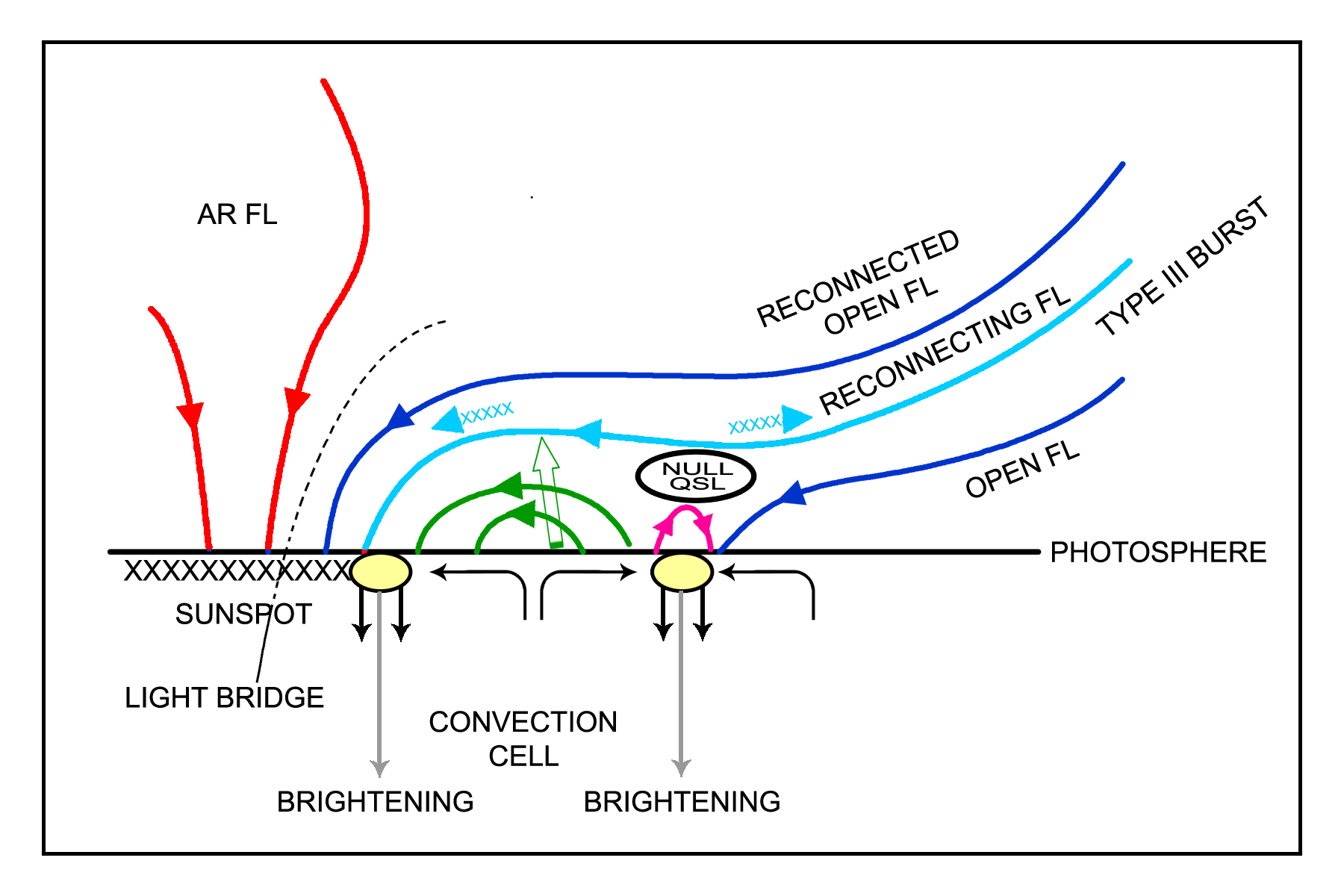}  
\caption{Schematic of topology of magnetic field at footpoint region of jet. The scenario explains the reconnection process linking the recurrent jets and the type~III radio burst.}\label{fig13_jet_model}
\end{center}
\end{figure}

Force-free extrapolations are equilibrium solutions and, therefore, do not allow for the direct modeling of the reconnection between field lines, which is an intrinsically dynamical process. However, the information about the magnetic field configuration discussed in Section~\ref{extrapolation} allows the depiction of a plausible scenario for the process yielding the jets and radio bursts, which is sketched in Fig.~\ref{fig13_jet_model}. 


Furthermore, the charted locations for $\omega=\omega_p$ on various open and closed field lines (see Fig.~\ref{fig7_radio_sources_imag}) presents numerously plausible locations for the radio sources. However, in reality, the radio sources suffer from propagation effects. The plasma distribution in the loop from the jet eruption may be responsible for enhancing density inhomogeneities. By comparing the $\omega=\omega_p$ locations on the loops with the observed radio source locations, we conclude that there is the presence of strong scattering in the apparent radio source locations, which are well-correlated in their temporal profiles. 

Next, we summarize a proposed physical scenario that is consistent
with both the observational data and the magnetic
models. Figure~\ref{fig13_jet_model} illustrates the proposed coronal
magnetic configuration of the region analyzed in this work. It is
surmised that a convection cell brought up field lines between the sunspot and the null (in green in Fig.~\ref{fig13_jet_model}), which were pushed toward the null point. Nulls are the preferential location where reconnection can occur, in particular, between the emerging flux and the ambient open field lines at the right hand side of the null  (the lower lying blue field line is on the right hand side in Fig.~\ref{fig13_jet_model}). Such open field lines were provided by a small coronal hole that was located on the southeast side of the active region. The product of the reconnection was two sets of field lines: one is represented in Fig.~\ref{fig13_jet_model} by the long field line that connects the periphery of the sunspot with the open field, the other is shown by the small pink loop below the null. Brightening would be expected at the footpoint of such reconnected field lines and it was indeed observed at the location of the footpoint of the field lines. On the other hand, the newly formed red field lines connecting to the open field were also energized by the reconnection, yielding the radio burst. Since the convection cell was steadily bringing up new flux, such a process could essentially be repeated in the same way over and over, yielding the homologous nature of the jets and radio bursts.

The multi-scale analysis of the field at local, AR, and solar-scales confirms the interlink between different flux bundles involved in the generation of the type~{\rm III} radio signal with flux transferred from a small coronal hole to the periphery of the sunspot via null point reconnection with an emerging structure. In this scenario, the sunspot light bridge did not have an active role in the process, but it was likely associated to the region of strong connectivity gradient separating that is open from AR field lines. PFSS extrapolations from higher resolution synoptic maps are necessary in order to check such a feature, which we reserve for future research.

\begin{acknowledgements}

This research was carried out when one of the authors, S.M.M. was a Ph.D. student at the University of Cambridge, UK. SMM acknowledges support from the Cambridge Trust, University of Cambridge, UK. At the start of this project, RS was a Ph.D. student at the National Centre for Radio Astrophysics (NCRA), Tata Institute of Fundamental Research, Pune. RS acknowledges the support of NCRA. RS appreciates and thanks to the Swiss National Foundation grant 200021\_175832 for their support. HEM, and GDZ acknowledge the support of STFC. AMV acknowledges ANPCyT grant PICT-2016/0221 and CONICET grant PIP-11220120100403 to IAFE that partially supported his participation in this research. GV acknowledges the support of the  Leverhulme Trust Research Project Grant 2014-05. \textrm{AIA} data are courtesy of SDO (NASA) and the \textrm{AIA} consortium. The HMI data used are courtesy of NASA/SDO and the HMI science team. This scientific work makes use of the Murchison Radio- astronomy Observatory, operated by CSIRO. We acknowledge the Wajarri Yamatji people as the traditional owners of the Observatory site. Support for the operation of the MWA is provided by the Australian Government Department of Industry and Science and Department of Education (National Collaborative Research Infrastructure Strategy: NCRIS), under a contract to Curtin University administered by Astronomy Australia Limited. We acknowledge the iVEC Petabyte Data Store and the Initiative in Innovative Computing and the CUDA Center for Excellence sponsored by NVIDIA at Harvard University. 

\end{acknowledgements}

\bibliographystyle{aa-note} 
\bibliography{jet_type3_paper.bib}    




\end{document}